\documentclass[a4paper,11pt]{article}
\usepackage{graphicx}
\usepackage[utf8]{inputenc}
\usepackage{amsfonts}
\usepackage{amssymb}
\usepackage{amsmath}
\usepackage{dsfont}
\usepackage{bbm}
\usepackage{framed}
\usepackage{fleqn}
\usepackage[colorlinks,pdfpagelabels,pdfstartview = FitH,bookmarksopen = true,bookmarksnumbered = true,linkcolor = black,plainpages = false,hypertexnames = false,citecolor = black]{hyperref}
\usepackage{longtable}
\usepackage{pdflscape}
\usepackage[ngerman,british]{babel}
\usepackage[tight,nice]{units}
\usepackage[width=16.5cm]{geometry}
\usepackage{color,cancel}
\usepackage[multiple]{footmisc} 
\usepackage[normalem]{ulem} 
\usepackage{url}

\definecolor{purple}{rgb}{0.5,0,0.5}

\newcommand{\ZZ}{\ensuremath{\mathbbm{Z}}}  
\newcommand{\QQ}{\ensuremath{\mathbbm{Q}}}  

\newcommand{\GL}[1]{\ensuremath{\mathrm{GL}(#1)}}  
\newcommand{\SL}[1]{\ensuremath{\mathrm{SL}(#1)}}
\newcommand{\PSL}[1]{\ensuremath{\mathrm{PSL}(#1)}}
\newcommand{\Dic}[1]{\ensuremath{\mathrm{Dic}_{#1}}}
\newcommand{\Carat}{{\sc carat}}

\newcommand{\bmat}{B}                      
\newcommand{\gram}{\mathrm{Gr}}            
\newcommand{\basis}[1]{\mathfrak{#1}}      
\newcommand{\scalarproduct}[2]{\left(#1,#2\right)} 
\newcommand{\form}[1]{\mathcal{#1}}        
\newcommand{\orth}{\ensuremath{\,\oplus\,}}        
\newcommand{\north}[1][]{\ensuremath{\,\odot_{#1}\,}} 
\newcommand{\arrowback}{\ensuremath{\hookleftarrow}} 
\newcommand{\emb}{\ensuremath{\hookrightarrow}}      
\newcommand{\mtwo}[1]{\left(\begin{array}{c@{\hspace{3mm}}c}#1\end{array}\right)}
\newcommand{\gtwo}[1]{$\mtwo{#1}$}
\newcommand{\mthree}[2]{\left(\begin{array}{c@{\hspace{#2 mm}}c@{\hspace{#2 mm}}c}#1\end{array}\right)}
\newcommand{\mthreer}[2]{\left(\begin{array}{r@{\hspace{#2 mm}}r@{\hspace{#2 mm}}r}#1\end{array}\right)}
\newcommand{\gthree}[2]{$\mthree{#1}{#2}$}

\definecolor{darkgreen}{rgb}{0.2,0.8,0.2}

\newcommand{\Apref}[1]{Appendix~\ref{#1}}
\newcommand{\Secref}[1]{Section~\ref{#1}}
\newcommand{\Eqref}[1]{Equation~\eqref{#1}}
\newcommand{\Figref}[1]{Figure~\ref{#1}}
\newcommand{\Tabref}[1]{Table~\ref{#1}}

\DeclareMathOperator{\tr}{tr}
\DeclareMathOperator{\Tr}{Tr}
\newcommand{\I}{\mathrm{i}}

\newcommand{\E}[1]{\ensuremath{\mathrm{E}_{#1}}} 

\newcommand{\SO}[1]{\ensuremath{\mathrm{SO}(#1)}}
\newcommand{\SU}[1]{\ensuremath{\mathrm{SU}(#1)}}

\newcommand{\Z}[1]{\ensuremath{\mathbbm{Z}_{#1}}} 

\newcommand{\ZxZ}[2]{\ensuremath{\mathbbm{Z}_{#1}\times\mathbbm{Z}_{#2}}}

\newcommand{\Id}[0]{\ensuremath{\mathbbm{1}}}
\newcommand{\GLabels}[2]{$\begin{array}{c}#1 \\ \left[#2\right] \end{array}$}
\newcommand{\GAP}[1]{$\left[#1\right]$}
\newcommand{\rep}[1]{\ensuremath\boldsymbol{#1}}
\newcommand{\crep}[1]{\ensuremath\bar{\boldsymbol{#1}}}

\numberwithin{equation}{section}
\numberwithin{table}{section}

\newcounter{dislist} 

\begin{document}

\begin{titlepage}

\vspace*{-3.0cm}
\begin{flushright}
\normalsize{DESY-12-147}\\
\normalsize{TUM-HEP 855/12}\\
\normalsize{FLAVOUR(267104)-ERC-29}\\
\normalsize{CETUP*-12/012}
\end{flushright}

\vspace*{1.0cm}

\begin{center}
{\Large\textbf{Classification of symmetric toroidal orbifolds}}

\vspace{1cm}

\textbf{
Maximilian~Fischer,
Michael~Ratz,
Jes\'us~Torrado
}
\\[5mm]
\textit{\small~Physik--Department T30, Technische Universit\"at M\"unchen,
James--Franck--Stra\ss e, 85748 Garching, Germany}
\\[8mm]
\textbf{
Patrick~K.S.~Vaudrevange
}
\\[5mm]
\textit{\small~Deutsches Elektronen--Synchrotron DESY, Notkestra\ss e 85, 22607 Hamburg, Germany}
\end{center}

\vspace{1cm}

\vspace*{1.0cm}

\begin{abstract}
We provide a complete classification of six--dimensional symmetric toroidal
orbifolds which  yield $\mathcal{N}\geq 1$ supersymmetry in 4D for the heterotic
string. Our strategy is based on a classification of crystallographic
space groups in six dimensions. We find in total 520 inequivalent  toroidal
orbifolds, 162 of them with Abelian point groups such as $\Z{3}$,  $\Z{4}$,
$\Z{6}$--I etc.\ and 358 with non--Abelian point groups such as $S_3$, $D_4$,
$A_4$ etc. We also briefly explore the properties of some orbifolds with Abelian
point groups and $\mathcal{N}=1$, i.e.\ specify the Hodge numbers and
comment on the possible mechanisms (local or non--local) of gauge 
symmetry breaking.
\end{abstract}
\end{titlepage}

\newpage
\tableofcontents
\newpage

\section{Introduction}

Heterotic string model building has received an increasing attention in the past
few years. The perhaps simplest heterotic compactifications are based on Abelian
toroidal orbifolds \cite{Dixon:1985jw,Dixon:1986jc}. Unlike in the supergravity
compactifications on Calabi--Yau manifolds one has a clear string theory
description.  In addition, the scheme is rich enough to produce a large number
of candidate models that may yield a stringy completion of the (supersymmetric)
standard model \cite{Ibanez:1987sn,Buchmuller:2005jr} (for a review see e.g.\
\cite{Nilles:2008gq}).
At the same time, symmetric orbifolds have a rather straightforward geometric
interpretation (cf.\ e.g.\ 
\cite{Kobayashi:1991rp,Forste:2004ie,Buchmuller:2006ik}). In fact, the geometric
properties often have immediate consequences for the phenomenological features
of the respective models. One obtains an intuitive understanding of discrete $R$
symmetries in terms of remnants of the Lorentz group of compact space, of the
appearance of matter as complete GUT multiplets due to localization properties
and gauge group topographies as well as flavor structures.

Despite their simplicity, symmetric toroidal orbifolds provide us with a
large number of different settings, which have, rather surprisingly, not been
fully explored up to now. In the past, different attempts of classifying 
(parts of) these compactifications have been made
\cite{Bailin:1999nk,Donagi:2008xy,Forste:2006wq,Dillies:2006yb}. These
classifications are not mutually consistent, and, as we shall see, incomplete.
The perhaps most complete classification is due to Donagi and Wendland (DW)
\cite{Donagi:2008xy}, who focus on $\ZxZ{2}{2}$ orbifolds. The main purpose of
this paper is to provide a complete classification of symmetric Abelian and
non--Abelian heterotic orbifolds that lead to $\mathcal{N}\ge1$ supersymmetry
(SUSY) in four dimensions.

The structure of this paper is as follows: in \Secref{sec:const} we  discuss the
tools used to construct toroidal orbifolds. Later, in  \Secref{sec:class}, we
present a way of classifying all possible  space groups that is novel in the
context of string compactifications. Then, in \Secref{sec:SUSY} we impose the
condition of $\mathcal{N}=1$ SUSY in 4D.  \Secref{sec:orbclass} is devoted to a
survey of the resulting orbifolds, and to a comparison with previous attempts to
classify Abelian symmetric orbifolds
\cite{Bailin:1999nk,Donagi:2008xy,Forste:2006wq,Dillies:2006yb}.  Finally, in
\Secref{sec:Discussion} we briefly discuss our results. In various appendices we
collect more detailed information on our classification program.
Appendix~\ref{app:details} contains some details on lattices, in Appendix 
\ref{app:2dimorb} we survey the already known 2D orbifolds, and in
Appendix~\ref{app:results} we provide tables of our results.

\section{Construction of toroidal orbifolds}
\label{sec:const}

We start our discussion with the construction of toroidal orbifolds 
\cite{Dixon:1985jw,Dixon:1986jc}. There are two equivalent ways of constructing 
such objects: (i) one can start from the Euclidean space $\mathbbm{R}^n$ and 
divide out a discrete group $S$, the so--called space group. (ii) 
Alternatively, one can start with an $n$--dimensional lattice $\Lambda$, 
to be defined in detail in \Secref{sec:lattice}, which determines a torus 
$\mathbbm{T}^n$ and divide out some discrete symmetry group $G$. Note that $G$, 
the so--called orbifolding group as defined in \Secref{sec:orbifoldinggroup}, 
is in general not equal to the point group introduced in 
\Secref{sec:pointgroups}. That is, a toroidal orbifold is defined as
\begin{equation}
\label{eqn:orbifold}
 \mathbbm{O} ~=~ \mathbbm{R}^n/S ~=~ \mathbbm{T}^n/G\;.
\end{equation}
Even though we are mostly interested in the case $n=6$ we will keep $n$ 
arbitrary. In the following, we will properly define the concepts behind 
\Eqref{eqn:orbifold}, closely following \cite{Brown:1978}.

\subsection[The space group $S$]{%
The space group $\boldsymbol{S}$}
\label{sec:spacegroup}

Let $S$ be a discrete subgroup of the group of motions in $\mathbbm{R}^n$, 
i.e.\ every element of $S$ leaves the metric of the space invariant. If
$S$ contains $n$ linearly independent translations, then it is called a
space group of degree $n$. Such groups appear already in crystallography:
they are the symmetry groups of crystal structures, which in turn
are objects whose symmetries comprise discrete translations.

Every element $g$ of a space group $S$ can be written as a composition of a 
mapping $\vartheta$ that leaves (at least) one point invariant and a 
translation by some vector $\lambda$, i.e.\ $g = \lambda \circ \vartheta$ 
for $g \in S$ (one can think of $\vartheta$ as a discrete rotation or 
inversion). This suggests to write a space group element as\footnote{In the 
mathematical literature the reverse notation $g=(\lambda,\vartheta)$ is also 
common, since the normal subgroup element is usually written to the left, and 
the lattice $\Lambda$ is the normal subgroup of the space group.}
\begin{equation}
g ~=~ (\vartheta, \lambda)\;,
\end{equation}
and it acts on a vector $v \in \mathbbm{R}^n$ as
\begin{equation}
v ~\stackrel{g}{\longmapsto}~ \vartheta\,v + \lambda\;.
\end{equation}
Let $h =(\omega, \tau) \in S$ be another space group element. Then the 
composition $h \circ g$ is given by $(\omega\,\vartheta, \omega\,\lambda +
\tau)$.

\subsection[The lattice $\Lambda$]{%
\texorpdfstring{The lattice $\boldsymbol{\Lambda}$}{The lattice Lambda}}
\label{sec:lattice}

Let $S$ be a space group. The subgroup $\Lambda$ of $S$ consisting of all 
translations in $S$ is the lattice of the space group. Note that for a 
general element $g = (\vartheta, \lambda) \in S$ the vector $\lambda$ does not 
need to be a lattice vector. Elements $g = (\vartheta, \lambda) 
\in S$ with $\lambda\notin\Lambda$ are called roto--translations.

Since a space group is required to contain $n$ linear independent translations, 
every lattice contains a basis $\basis{e}=\{e_i\}_{i\in\{1,\ldots,n\}}$ and the 
full lattice is spanned by the $e_i$ (with integer coefficients),
i.e.\ an element  $\lambda \in \Lambda$ can be written as $\lambda = n_i\, e_i$,
summing over  $i=1,\ldots,n$ and $n_i \in \ZZ$. Clearly, the choice of basis is
not unique.  For example, for a given lattice $\Lambda$ take two bases 
$\basis{e}=\{e_1,\ldots,e_n\}$ and $\basis{f}=\{f_1,\ldots,f_n\}$ and  define
$\bmat_{\basis{e}}$ and $\bmat_{\basis{f}}$ as matrices whose columns  are the
basis vectors in $\basis{e}$ and $\basis{f}$, respectively. Then the  change of
basis is given by a unimodular matrix $M$ (i.e.\ $M\in\GL{n,\ZZ}$) as 
\begin{equation}\label{eq:changeb}
 \bmat_{\basis{e}}\, M ~=~ \bmat_{\basis{f}}\;.
\end{equation}
On the other hand, one can decide whether two bases 
$\basis{e}$ and $\basis{f}$ span the same lattice by computing the 
matrix $M~=~\bmat_\basis{e}^{-1}\,\bmat_\basis{f}$ and checking whether or not 
it is an element of $\GL{n,\ZZ}$.

\subsection[The point group $P$]{%
\texorpdfstring{The point group $\boldsymbol{P}$}{The point group P}}
\label{sec:pointgroups}

For a space group $S$ with elements of the form $(\vartheta, \lambda)$, the set 
$P$ of all $\vartheta$ forms a finite group (\cite[p.~15]{Brown:1978}), the 
so--called point group of $S$. The elements of a point group are sometimes 
called twists or rotations. However, in general a point group can also contain 
inversions and reflections, i.e.\ $\vartheta\in\mathrm{O}(n)$.

The point group $P$ of $S$ maps the lattice of $S$ to itself. Hence, similarly
to  the change of lattice bases, point group elements can be represented by
$\GL{n,\ZZ}$ (i.e.\ unimodular) matrices. When written in  the $\GL{n,\ZZ}$
basis, we append  the twists by an index indicating the lattice basis, while the
$\mathrm{O}(n)$ (or \SO{n}) representation of the twist is denoted without an
index. For example,  the twist $\vartheta\in\mathrm{O}(n)$ is denoted as
$\vartheta_\basis{e}$ in  the lattice basis $\basis{e}=\{e_1,\ldots,e_n\}$ such
that $\vartheta\, e_i = \left(\vartheta_\basis{e}\right)_{ji}\, e_j$ 
and $\vartheta_\basis{e} = \bmat_\basis{e}^{-1}\, \vartheta\, \bmat_\basis{e}$. 
Furthermore, under a change of basis as in \Eqref{eq:changeb} the twist 
transforms according to
\begin{equation}
\label{eqn:BasisChangeTheta}
\vartheta_{\basis{f}} ~=~ M^{-1}\,\vartheta_{\basis{e}}\,M\;.
\end{equation}

Given these definitions, and provided that the lattice is always a normal 
subgroup of the space group (i.e.\ rotation $\circ$ translation $\circ$
(rotation)$^{-1}$ $=$ translation), the space group $S$ has a semi--direct
product structure iff the point group $P$ is a subgroup of it, i.e.\
$P\subset S$. In that case
\begin{equation}
S~=~P\ltimes \Lambda\;,
\end{equation}
and one can write the orbifold as
\begin{equation}
\mathbbm{O} ~=~ \mathbbm{R}^n/(P\ltimes\Lambda) ~=~ \mathbbm{T}^n/P\;.
\end{equation}

In general, however, the point group is not a subgroup of the space group and 
thus the space group is not necessarily a semi--direct product of its point 
group with its lattice. More precisely, in general the point group $P$ is not 
equal to the orbifolding group $G$ of \Eqref{eqn:orbifold} because of the
possible presence of roto--translations, as we will see in an example in 
\Secref{sec:spacegroupexamples}.

\subsection[Examples: space groups with $\Z{2}$ point group]{%
\texorpdfstring{Examples: space groups with $\boldsymbol{\Z{2}}$ point group}{Examples: space groups with Z2 point group}}
\label{sec:spacegroupexamples}

In this section, we give two examples of space groups in two dimensions with 
$\Z{2}$ point group in order to illustrate the discussion of the previous 
sections.

\begin{figure}[ht]
\centering
\begin{tabular}{c@{\hspace{20mm}}c}
(a)\, \includegraphics[height=25mm]{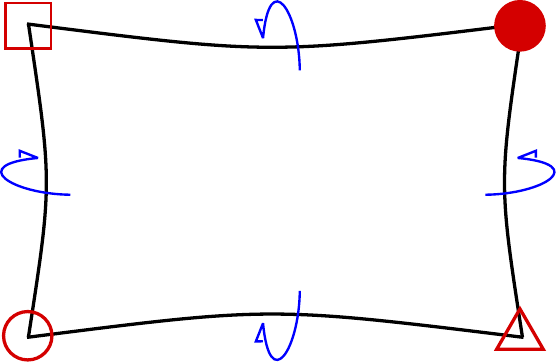}&
(b)\, \includegraphics[height=25mm]{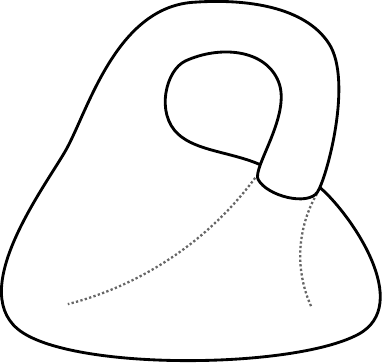}
\end{tabular}
\caption{Two--dimensional examples: (a) ``pillow'' and (b) Klein bottle. In 
case (a) the blue arrows indicate a wrap--around and the red symbols indicate 
fixed points.}
\label{fig:examples}
\end{figure}

\subsubsection*{A simple example: the ``pillow''}

The first of our examples is the well known two--dimensional ``pillow'', see  
\Figref{fig:examples}(a). The space group $S$ is generated as
\begin{equation}
S ~=~ \left\langle(\mathbbm{1},e_1),(\mathbbm{1},e_2),(\vartheta,0)\right\rangle\;,
\end{equation}
and can be realized as the semi--direct product of the oblique lattice 
$\Lambda$ (see \Apref{app:bravaislie}) and the point group 
$P = \{\mathbbm{1}, \vartheta\}$. In detail, the lattice is given as 
$\Lambda = \{ n_1\, e_1 + n_2\, e_2, n_i \in\Z{} \}$ using the basis 
$\basis{e} = \{e_{1},e_{2}\}$. $\vartheta$ is a rotation by $\pi$, i.e.\ it 
acts on the lattice basis vectors as
\begin{equation}
\vartheta\ e_i ~=~ -e_i \qquad \text{for}\qquad i~=~1,2\;.
\end{equation}
Therefore, it can be represented by a $\GL{2,\ZZ}$ matrix
\begin{equation}
\vartheta_\basis{e} ~=~ \left(\begin{array}{cc}-1&0\\0&-1\end{array}\right)\;.
\end{equation}
Since $\vartheta^2 = \mathbbm{1}$, the point group is $\ZZ_2$.

\subsubsection*{Another example: the Klein bottle}

Let us take a look at a more advanced example: the space group of a Klein bottle, 
see \Figref{fig:examples}(b). Here, the space group is generated by two 
orthogonal lattice vectors (which thus span a primitive rectangular lattice $\Lambda$) 
$\{e_1,e_2\}$, and an additional element $g$,
\begin{equation}
S ~=~ \left\langle(\mathbbm{1},e_1), (\mathbbm{1},e_2), g\right\rangle
\qquad\text{with}\qquad
g ~=~ \left(\vartheta,\tfrac{1}{2}e_1\right)
\qquad\text{and}\qquad
\vartheta_\basis{e} ~=~ \left(\begin{array}{cc}1&0\\0&-1\end{array}\right)\;.
\end{equation}
$g$ acts on a vector $v=v^1 e_1+v^2 e_2$ as
\begin{equation}
v ~\stackrel{g}{\longmapsto}~ \vartheta\, v + \frac{1}{2}e_1 
  ~=~ v^1\, e_1 - v^2\, e_2 + \frac{1}{2}e_1\;.
\end{equation}
Notice that even though the point group is $\ZZ_2$ (i.e.\ 
$\vartheta^2 = \mathbbm{1}$), $g$ generates a finite group isomorphic to 
$\mathbbm{Z}_2$ only on the torus $\mathbbm{T}^2~=~\mathbbm{R}^2/\Lambda$, 
but not on the Euclidean space $\mathbbm{R}^2$, because $g^2 = 
(\mathbbm{1},e_1) \neq (\Id,0)$. In other words, since the generator $g$ also 
contains a translation $\frac{1}{2}e_1 \notin \Lambda$, it is not a point group 
element but a roto--translation.

Obviously, this space group cannot be written as a semi--direct product of a 
lattice and a point group, as is always the case when we have 
roto--translations.

\subsection[The orbifolding group $G$]{%
\texorpdfstring{The orbifolding group $\boldsymbol{G}$}{The orbifolding group G}}
\label{sec:orbifoldinggroup}

Due to the possible presence of roto--translations, it is clear that in general 
space groups cannot be described by lattices and point groups only. Therefore, 
we will need to define an additional object, the orbifolding group (see 
\cite{Donagi:2008xy}). Loosely speaking, the orbifolding group $G$ is generated 
by those elements of $S$ that have a non--trivial twist part, identifying
elements which differ by a lattice translation. Hence, if there 
are no roto--translations the orbifolding group $G$ is equal to the point group 
$P$. In other words, the orbifolding group may contain space group elements 
with non--trivial, non--lattice translational parts. Combining the orbifolding 
group $G$ and the torus lattice $\Lambda$ generates the space group $S = 
\langle\{ G,\Lambda \}\rangle$. 

Hence, we can define the orbifold as
\begin{equation} \mathbbm{O} ~=~ \mathbbm{R}^n/S ~=~ \mathbbm{R}^n/\langle\{G,\Lambda\}\rangle ~=~ (\mathbbm{R}^n/\Lambda)/G ~=~ \mathbbm{T}^n/G\;.
\end{equation}
Orbifolds can be manifolds (see e.g.\ \Figref{fig:examples}(b)), but in 
general, they come with singularities which can not be endowed 
with smooth maps (see e.g.\ \Figref{fig:examples}(a)).

\section{Equivalences of space groups}
\label{sec:class}

\begin{figure}[t]
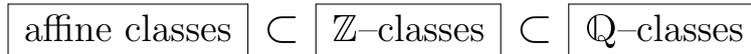

\renewcommand{\arraystretch}{1.5}
\centering
\begin{tabular}{|c|c|c|c|c|}
\cline{1-1}\cline{3-3}\cline{5-5}
{\Large affine classes}
&{\LARGE $\subset$}
&{\Large \ZZ--classes}
&{\LARGE $\subset$}
&{\Large \QQ--classes}\\
\cline{1-1}\cline{3-3}\cline{5-5}
\end{tabular}
\caption{Sketch of the classification of space groups.}
\label{fig:class}
\end{figure}

In the context of string orbifold compactifications, some physical properties 
of a given model directly depend on the choice of its space group. These 
features are common to whole sets of space groups and can be related  to
some mathematical properties. Using the latter, one can 
define equivalence classes of space groups. In detail, there are three kinds of 
equivalence classes suitable to sort space groups $S$ with certain physical and 
corresponding mathematical properties. These classes are:
\begin{enumerate}
\item the \QQ--class (see \Secref{sec:Qclass}) determines the point group
$P$ contained in $S$ and hence the number of supersymmetries in 4D and the
number of geometrical moduli;
\item the \ZZ--class (see \Secref{sec:Zclass}) determines the lattice
$\Lambda$ of $S$ and hence the nature of the geometrical moduli;
\item the affine class (see \Secref{sec:affineclass}) determines the
flavor group and the nature of gauge symmetry breaking (i.e.\ local vs.\  
non--local gauge symmetry breaking).
\end{enumerate}
Each \QQ--class can contain several \ZZ--classes and each \ZZ--class can contain 
several affine classes, see \Figref{fig:class}. In other words, for every point 
group there can be several inequivalent lattices and for every lattice there 
can be several inequivalent choices for the orbifolding group (i.e.\ with or 
without roto--translations).

In the following, we will discuss in detail why the concept of affine classes  
is advantageous to classify physically inequivalent space groups. This is
standard knowledge among crystallographers and can for instance be found in
more detail in \cite{Brown:1978}.

\subsection{Affine classes of space groups}
\label{sec:affineclass}

Two space groups $S_1$ and $S_2$ of degree $n$ belong to the same affine class 
(i.e.\ $S_1 \sim S_2$) if there is an affine mapping
$f~:~\mathbbm{R}^n\to\mathbbm{R}^n$ such that
\begin{equation}\label{eq:EquivalenceOfSpaceGroups}
 f^{-1}\,S_1\,f ~=~ S_2\;.
\end{equation}
An affine mapping $f = (A,t)$ on $\mathbbm{R}^n$ consists of a translation $t$ 
and a linear mapping $A$, that is, it allows for rescalings and rotations. 
Therefore, this definition enables us to distinguish between space groups that 
actually describe different symmetries and space groups which are just the ones 
we already know, looked upon from a different angle or distance. Then, for a 
given representative space group of an affine class a non--trivial affine
transformation $A$ that leaves the point group invariant (i.e.\
$A^{-1}\,P\,A~=~P$)  corresponds to a change of the geometrical data. In the
context of superstring compactifications this corresponds to a change of values
of the geometrical moduli. That is, affine transformations amount to moving in
the moduli space of the respective compactification. Hence, we will  only be
interested in one representative for every affine class. 

It turns out that, for a given dimension $n$, there exists only a
finite number of affine classes of space groups \cite[p.~10]{Brown:1978}.
Hence, classifying all affine classes of space groups enables a
complete classification of orbifolds for a fixed number of dimensions. In
this paper, we focus on the six--dimensional case.

\subsubsection*{Example in two dimensions}

Let us illustrate this at the $\mathbbm{T}^2/\Z{2}$ example with 
$\vartheta = -\Id$ given in \Secref{sec:spacegroupexamples}. As 
discussed there, the lattice is oblique, i.e.\ one can choose any linear 
independent vectors $e_1$ and $e_2$ as basis vectors. Define a space group $S$ 
by choosing
\begin{equation}
e_1 ~=~ \left(\begin{array}{c} r_1\\0\end{array}\right) \quad\text{and}\quad
e_2 ~=~ \left(\begin{array}{c} r_2 \cos(\alpha)\\r_2 \sin(\alpha)\end{array}\right)\;.
\end{equation}
This space group is in the same affine class as $\widetilde{S}$ with basis vectors
\begin{equation}
\widetilde{e}_1 ~=~ \left(\begin{array}{c} 1\\0\end{array}\right) \quad\text{and}\quad
\widetilde{e}_2 ~=~ \left(\begin{array}{c} 0\\1 \end{array}\right)\;.
\end{equation}
This can be seen explicitly using the affine transformation 
$f = (A,0)$ with
\begin{equation}
A ~=~ \left(\begin{array}{cc} r_1 & r_2 \cos(\alpha)\\ 0 & r_2 \sin(\alpha)\end{array}\right) 
\quad\text{and}\quad
A^{-1} ~=~ \left(\begin{array}{cc} \frac{1}{r_1} & -\frac{1}{r_1\tan(\alpha)}\\ 0 & \frac{1}{r_2 \sin(\alpha)}\end{array}\right)\;.
\end{equation}
Take an arbitrary element $g = (\vartheta,n_i e_i)$ with 
$n_i\in\ZZ$ for $i=1,2$. Then
\begin{subequations}
\begin{eqnarray}
\left(f^{-1}\,g\,f\right)(x) & = & \left(f^{-1}\,g\right)(A x) ~=~ f^{-1}(\vartheta A x + n_i e_i) ~=~ \vartheta x + A^{-1}(n_i e_i)\\
                             & = & \vartheta x + n_i\, \widetilde{e}_i ~=~ \widetilde{g}\, x
\end{eqnarray}
\end{subequations}
for $x \in \mathbbm{R}^2$ and $\widetilde{g}=(\vartheta, n_i \widetilde{e}_i)\in\widetilde{S}$. 
Therefore, $S \sim \widetilde{S}$ and there is only one affine class of
$\mathbbm{T}^2/\Z{2}$ space groups with $\vartheta = -\Id$.

This should be compared with the $\mathbbm{T}^2/\Z3$ orbifold, where the
angle between the basis vectors $e_i$ and their length ratio are fixed, such
that the corresponding moduli space is different. Hence, it is clear that
$\mathbbm{T}^2/\Z2$ and $\mathbbm{T}^2/\Z3$ are two different orbifolds.
This demonstrates the advantages of using affine classes for the classification of 
space groups.

\subsection[$\ZZ$--classes of space groups]{%
\texorpdfstring{$\boldsymbol{\ZZ}$--classes of space groups}{Z-classes of space groups}}
\label{sec:Zclass}

As discussed above, we can sort space groups into affine classes. This can
be refined further by grouping affine classes according to common properties
of their point groups. Following the argument in \Secref{sec:pointgroups}, the 
elements of the point group can be written in the lattice basis as elements of 
$\GL{n,\ZZ}$. Therefore, a point group is a finite subgroup of the unimodular 
group on \ZZ.

Take two space groups $S_1$ and $S_2$. For $i=1,2$, the space group $S_i$ 
contains a lattice $\Lambda_i$ and its point group in the lattice basis is 
denoted by $P_i$, i.e.\ $P_i \subset \GL{n,\ZZ}$. Then, the two space groups 
belong to the same \ZZ--class (or in other words to the same arithmetic crystal 
class) if there exists an unimodular matrix $U$ (i.e.\ $U\in\GL{n,\ZZ}$) such 
that (cf.\ the parallel discussion around
\Eqref{eq:EquivalenceOfSpaceGroups})
\begin{equation}
\label{eqn:Zclass}
U^{-1}\, P_1\, U ~=~ P_2\;,
\end{equation}
see \Eqref{eqn:BasisChangeTheta}. That is, if the point groups are related by a 
change of lattice basis (using $U$), the space groups belong to the same 
\ZZ--class. Hence, \ZZ--classes classify the inequivalent lattices.

If two space groups belong to the same \ZZ--class, they have the same form
space  and, physically, they possess the same amount and nature of geometrical
moduli.  However, as we have stressed before, space groups from the same
\ZZ--class are  not necessarily equivalent because of the possible presence
of roto--translations. In other words, space  groups from the same \ZZ--class
can belong to different affine classes and  can hence be inequivalent.

\subsection[$\QQ$--classes of space groups]{%
\texorpdfstring{$\boldsymbol{\QQ}$--classes of space groups}{Q-classes of space groups}}
\label{sec:Qclass}

As before in \Secref{sec:Zclass}, take two space groups $S_1$ and $S_2$. 
For $i=1,2$, the point group in the lattice basis associated to the space group 
$S_i$ is denoted by $P_i$, i.e.\ $P_i \subset \GL{n,\ZZ}$. Then, the two space 
groups belong to the same \QQ--class (or in other words to the same geometric 
crystal class) if there exists a matrix $V\in\GL{n,\QQ}$ such that
\begin{equation}
\label{eqn:Qclass}
 V^{-1}\, P_1\, V ~=~ P_2\;.
\end{equation}
Obviously, if two space groups belong to the same \ZZ--class they also belong 
to the same \QQ--class, hence the inclusion sketch in \Figref{fig:class}. In 
contrast to \ZZ--classes, \QQ--classes do not distinguish between  inequivalent
lattices. However, if two space groups belong to the same  \QQ--class, the
commutation relations and the orders of the corresponding point  groups are the
same. Therefore, they are isomorphic as crystallographic point  groups. They
also possess form spaces of the same dimension, i.e.\ they  have the same number
of moduli. What is important for physics is that all space groups in the same 
\QQ--class share a common holonomy group (cf.\ \Secref{sec:SUSY}). This allows 
us to identify settings that yield $\mathcal{N}=1$ SUSY in 4D. In particular, 
in order to determine the number of SUSY generators, it is sufficient to 
consider only one  representative from every \QQ--class.

\subsection{Some examples}
\label{subsec:class:ex}

Before going to six dimensions, let us illustrate the above definitions with 
some easy examples of two--dimensional $\Z{2}$ orbifolds, taken from 
\Apref{app:2dimorb}.

\subsubsection*{Space groups in the same $\boldsymbol{\ZZ}$--class}

\begin{figure}
\centering
\includegraphics[height=38mm]{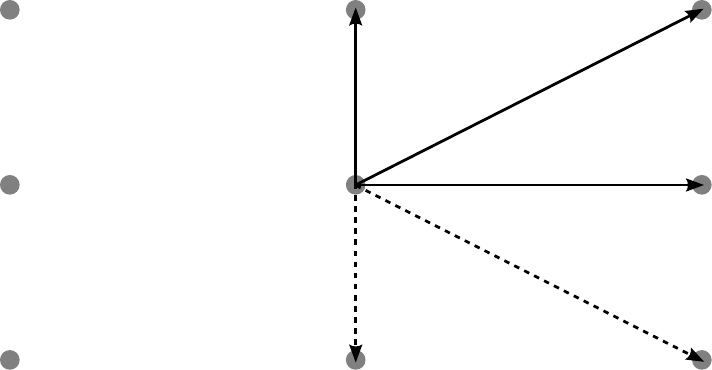}
\put(2,36){$f_2$}
\put(2,18){$e_1\equiv f_1\equiv e_1^\prime\equiv f_1^\prime$}
\put(2,0){$f_2^\prime$}
\put(-34,36){$e_2$}
\put(-34,0){$e_2^\prime$}
\caption{Two different bases for the p--rectangular lattice: $\basis{e}=\{e_1,e_2\}$ 
and $\basis{f}=\{f_1,f_2\}$, and the action of the point group generator 
(primed vectors).}
\label{fig:bases}
\end{figure}
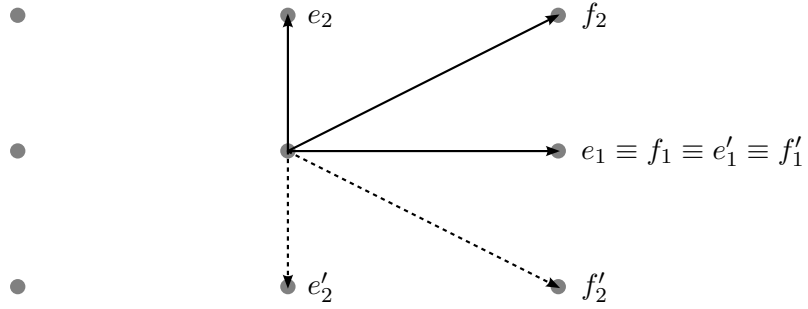

Consider the affine class \Z{2}--II--1--1, as defined in 
\Apref{app:2dimorb}. As there are no roto--translations, the orbifolding group 
is equal to the point group and is generated by $\vartheta$, a reflection at 
the horizontal axis. Now, let this reflection act on a lattice, first spanned 
by the basis vectors $\basis{e}=\{e_1,e_2\}$ and second spanned by 
$\basis{f}=\{f_1,f_2\}$, see \Figref{fig:bases}. The two corresponding space 
groups read
\begin{eqnarray}
S_\basis{e}&=&\langle(\vartheta,0),(\mathds{1},e_1),(\mathds{1},e_2)\rangle
\qquad\text{with}\qquad\vartheta_\basis{e}~=~\left(\begin{array}{cc}1&0\\0&-1\end{array}\right)\;,\\
S_\basis{f}&=&\langle(\vartheta,0),(\mathds{1},f_1),(\mathds{1},f_2)\rangle
\qquad\text{with}\qquad\vartheta_\basis{f}~=~\left(\begin{array}{cc}1&2\\0&-1\end{array}\right)\;,
\end{eqnarray}
where $\vartheta_\basis{e} \neq \vartheta_\basis{f}$ because they are given in their 
corresponding lattice bases. However, it is easy to see that they are related by
the $\GL{2,\ZZ}$ transformation
\begin{equation}
U~=~\left(\begin{array}{cc}1&1\\0&1\end{array}\right) 
\quad\text{with}\quad U^{-1}\, \vartheta_\basis{e}\, U ~=~ \vartheta_\basis{f}\;,
\end{equation}
cf.\ \Eqref{eqn:Zclass}. Therefore, they belong to the same \ZZ--class. Hence, as 
we actually knew from the start, they act on the same lattice and the matrix 
$U$ just defines the associated change of basis precisely as in 
\Eqref{eq:changeb}.

\subsubsection*{Space groups in the same $\boldsymbol{\QQ}$--class, but different $\boldsymbol{\ZZ}$--classes}

Next, consider the space groups,
\begin{eqnarray}
S_\text{1--1}&=&\langle(\vartheta_\text{1--1},0),(\mathds{1},e_1),(\mathds{1},e_2)\rangle
\qquad\text{with}\qquad\vartheta_{\text{1--1},\basis{e}} ~=~ \left(\begin{array}{cc}1&0\\0&-1\end{array}\right)\;,\\
S_\text{2--1}&=&\langle(\vartheta_\text{2--1},0),(\mathds{1},f_1),(\mathds{1},f_2)\rangle
\qquad\text{with}\qquad\vartheta_{\text{2--1},\basis{f}} ~=~ \left(\begin{array}{cc}0&1\\1&0\end{array}\right)\;,
\end{eqnarray}
with lattices spanned by $e_1=(1,0)$, $e_2=(0,1)$ and 
$f_1=(\nicefrac{1}{2},\nicefrac{1}{2})$,
$f_2=(\nicefrac{1}{2},-\nicefrac{1}{2})$, respectively.
The first space group belongs to the affine class \Z{2}--II--1--1 and 
the second one to \Z{2}--II--2--1, see \Apref{app:2dimorb}. If we try 
to find the transformation $V$ from \Eqref{eqn:Qclass} that fulfills 
$V^{-1}\, \vartheta_{\text{1--1},\basis{e}}\, V = \vartheta_{\text{2--1},\basis{f}}$ 
we see that
\begin{equation}\label{eq:mdiffz}
V~=~\left(\begin{array}{cc}x&x\\y&-y\end{array}\right)\qquad\text{with}\qquad x,y~\in~\QQ\;.
\end{equation}
But for all values of $x$ and $y$ for which $V^{-1}$ exists, either $V$ or 
$V^{-1}$ has non--integer entries. Therefore, the space groups 
\Z{2}--II--1--1 and \Z{2}--II--2--1 belong to the same 
\QQ--class, but to different \ZZ--classes. In other words, these space groups 
are defined with inequivalent lattices. Indeed, the first space group possesses 
a primitive rectangular lattice, while the second one has a centered 
rectangular lattice, as we will see in detail in the following.

\subsubsection*{The effect of including additional translations}

There is an alternative way of seeing the relationship between the two space 
groups of the last example: one can amend one of the  space groups by an
additional translation. In general, this gives rise to a new lattice, and
consequently to a different \ZZ--class.

\begin{figure}[ht]
\begin{center}
\includegraphics[height=38mm]{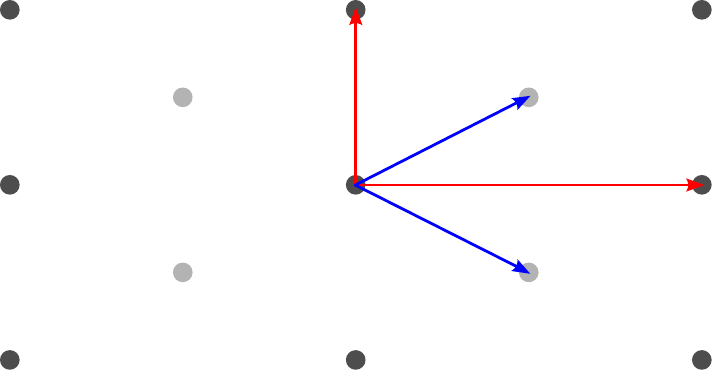}
\put(-34,36){$e_2$}
\put(2,18){$e_1$}
\put(-16,27.5){$\tau$}
\put(-16,9){$e_1-\tau$}
\caption{Change of a lattice by an additional translation: the basis of the 
original lattice is red, the basis of the new one blue. The additional lattice points 
are gray. The action of $\vartheta$ is a reflection at the horizontal axis. 
Therefore, it maps $e_1$ to itself, $e_2$ to its negative and interchanges 
$\tau$ and $e_1-\tau$.}
\label{fig:addshift}
\end{center}
\end{figure}
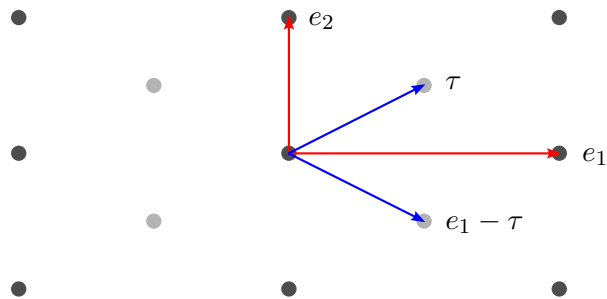

In our case, let us take the \Z{2}--II--1--1 affine class and add the 
non--lattice translation 
\begin{equation}
\tau~=~\frac{1}{2}(e_1+e_2)
\end{equation}
to its space group. If we incorporate this translation into the lattice, we 
notice that this element changes the original primitive rectangular lattice to 
a centered rectangular lattice, with a unit cell of half area. The new 
lattice (see \Figref{fig:addshift}) can be spanned by the basis vectors $\tau$ 
and $e_1-\tau$. 

We can interpret the inclusion of this additional translation as a ``change of 
basis'', see \Eqref{eq:changeb}, but now generated by a matrix 
$M\in\GL{2,\QQ}$ instead of one from $\GL{2,\ZZ}$. The transformation looks 
like
\begin{equation}
\bmat_{\basis{e}}\, M ~=~ \bmat_{\basis{\tau}}\quad\text{with}\quad
M~=~\left(\begin{array}{cc}\nicefrac{1}{2}&\nicefrac{1}{2}\\\nicefrac{1}{2}&-\nicefrac{1}{2}\end{array}\right)\;,
\end{equation}
where $\bmat_{\basis{e}}$ and $\bmat_{\basis{\tau}}$ are matrices whose columns 
are $(e_1, e_2)$ and $(\tau, e_1-\tau)$, respectively. $M$ is precisely the 
matrix in \Eqref{eq:mdiffz} with values $x=y=\nicefrac{1}{2}$. Performing this 
basis change, the twist has to be transformed accordingly. Hence, the two 
\ZZ--classes are related by a $\GL{2,\QQ}$ transformation $M$ and the new space 
group with lattice $\bmat_{\basis{\tau}}$ is \Z{2}--II--2--1. The 
geometrical action of the twist, however, is the same in both cases: it is a 
reflection at the horizontal axis (see \Figref{fig:addshift}). That is the 
reason for the name geometrical crystal classes for \QQ--classes. A general 
method for including additional translations can be found in 
\Apref{app:latt:add}.

The method of using additional translations has been used in 
\cite{Donagi:2008xy} and \cite{Dillies:2006yb} in order to classify 
six--dimensional space groups with point groups $\ZxZ{N}{N}$ for 
$N=2,3,4,6$ (the classification of \cite{Dillies:2006yb} is not fully
exhaustive). 
In these works, the authors start with factorized lattices, i.e.\ lattices
which are the  orthogonal sum of three two--dimensional sublattices, on which
the twists act  diagonally. Then, in a second step additional translations are
introduced.  As we have shown here, adding such translations is equivalent to
switching  between \ZZ--classes in the same \QQ--class. Hence, if one considers
all  possible lattices (\ZZ--classes) additional translations do not give rise
to  new orbifolds.

\subsubsection*{Space groups in different $\boldsymbol{\QQ}$--classes}

Finally, consider the affine classes \Z{2}--I--1--1 and 
\Z{2}--II--1--1 defined in \Apref{app:2dimorb}. If we try to
find a transformation between both space groups generators, see
\Eqref{eqn:Qclass},
\begin{equation}
V^{-1}\,\left(\begin{array}{cc}-1&0\\0&-1\end{array}\right)\,V 
~=~ \left(\begin{array}{cc}1&0\\0&-1\end{array}\right) \;\Leftrightarrow\;
\left(\begin{array}{cc}-1&0\\0&-1\end{array}\right)\,V 
~=~ V\,\left(\begin{array}{cc}1&0\\0&-1\end{array}\right)\;,
\end{equation}
we obtain
\begin{equation}
V~=~\left(\begin{array}{cc}0&x\\0&y\end{array}\right)\; ~\notin~\GL{2,\QQ}\qquad \forall\, x,y\;.
\end{equation}
Therefore, the space groups \Z{2}--I--1--1 and 
\Z{2}--II--1--1 belong to different \QQ--classes (and also to 
different \ZZ--classes). That is, the point groups are inequivalent: the twist 
of the first point group is a reflection at the origin and the twist of the 
second point group is a reflection at the horizontal axis.

\section{Classification of space groups}
\label{sec:SUSY}

In this section we describe our strategy to classify all inequivalent space 
groups for the compactification of the heterotic string to four dimensions with 
$\mathcal{N}=1$ SUSY.

\subsection{Classification strategy}

As is well known, the amount of residual supersymmetry exhibited by the 4D
effective theory is related to the holonomy group of the compact space
\cite{Candelas:1985en}. In the context of orbifolds, one can relate the holonomy
group to the point group \cite{Dixon:1986jc}.
Orbifold compactifications preserve four--dimensional supersymmetry if the 
point group is a discrete subgroup of $\SU{3}$. Hence, the amount of unbroken 
SUSY is the same for all members of a given \QQ--class. Therefore, we start our 
classification with the identification of all \QQ--classes (i.e.\ point groups) 
that are subgroups of $\SU{3}$. Then, for each \QQ--class we identify all 
\ZZ--classes (i.e.\ lattices) and finally construct for each \ZZ--class all 
affine classes (i.e.\ roto--translations).

In more detail, our strategy reads:
\begin{enumerate}
\item Choose a \QQ--class and find a representative $P$ of it.\footnote{A 
discussion about the possible orders of the elements of the point group, and 
therefore the possible point groups, can be found in \Apref{app:2dimorb}.}
\item Check that $P$ is a subgroup of $\SO{6}$ rather than $\mathrm{O}(6)$.
\item Verify that $P$ is a subgroup of $\SU{3}$.
\item Find every possible \ZZ--class inside that \QQ--class.
\item Find every possible affine class inside each one of those \ZZ--classes. 
\end{enumerate}

There exists a catalog of every possible affine class in up to six dimensions 
classified into \ZZ-- and \QQ--classes \cite{Plesken:1998}. Furthermore, one can 
access this catalog easily using the software \Carat~\cite{CARAT}.
In detail, the command 
\texttt{Q\_catalog} lists all \QQ--classes, the command \texttt{QtoZ} lists all 
\ZZ--classes of a given \QQ--class and, finally, the command \texttt{Extensions} 
lists all affine classes of a given \ZZ--class. Hence, the main open question 
is to decide whether a given representative of a \QQ--class is a subgroup of 
$\SU{3}$.

\subsection{Residual SUSY}

We start by verifying that $P \subset \SO{6}$. \Carat\ offers representatives for 
all \QQ--classes, i.e.\ it gives the generators of the point group $P$ in some 
(unspecified) lattice basis $\basis{e}$ as $\GL{6,\ZZ}$ matrices 
$\vartheta_\basis{e}$. In principle, one can transform them to matrices from 
$\mathrm{O}(6)$ using the (unspecified) lattice basis, i.e.\ 
$\vartheta ~=~ B_\basis{e} \,\vartheta_\basis{e}\, B_\basis{e}^{-1}$. However, 
as the determinant is invariant under this transformation 
($\text{det}(\vartheta) = \text{det}(\vartheta_\basis{e})$) one can check
whether or not the determinant equals $+1$ for all generators of $P$ in the 
$\GL{6,\ZZ}$ form given by \Carat. This allows us to determine whether or not 
$P \subset \SO{6}$.

Next, we recall that the matrices $\vartheta_\basis{e} \in P$ originate from the 
six--dimensional representation $\rep{6}$ of $\SO{6}$. One way to check that 
$P$ is a subgroup of $\SU{3}$ is to consider the breaking of the $\rep{6}$ into 
representations of $\SU{3}$,
\begin{equation}
\rep{6} ~\rightarrow~ \rep{3} \oplus \crep{3}\;.
\end{equation}
On the other hand, the six--dimensional representation is, in general, a
reducible representation of the point group $P$. Hence, it can be decomposed
\begin{equation}
\label{eqn:decomposeintoPreps}
\rep{6} ~\rightarrow~ \rep{a} \oplus \rep{b} \oplus \ldots
\end{equation}
into irreducible representations $\rep{a}, \rep{b}, \ldots$ of $P$. This 
decomposition can be computed using the character table of $P$ as discussed in 
the following. 

For $g\in P$, the character $\chi_{\rep{\rho}}(g)$ in the representation 
$\rep{\rho}$ is given by the trace of the matrix representation $\rep{\rho}(g)$ 
of $g$,
\begin{equation}
\chi_{\rep{\rho}}(g) ~=~ \Tr(\rep{\rho}(g))\;.
\end{equation}
As the trace is invariant under cyclic permutations, the character 
$\chi_{\rep{\rho}}$ is the same for all elements of a conjugacy class, i.e.\
\begin{equation}
\chi_{\rep{\rho}}(g) ~=~ \chi_{\rep{\rho}}(h) \quad\text{for}\quad h ~\in~ \left[g\right] ~=~ \{ f\,g\,f^{-1} \text{ for all } f \in P\}\;.
\end{equation}
Now, the character table of a finite group $P$ contains one row for each 
irreducible representation $\rep{\rho}_i$ and one column for each conjugacy 
class $\left[g_j\right]$ and the entry is the corresponding character 
$\chi_{\rep{\rho}_i}(g_j)$. In fact, the number of irreducible 
representations $c$ equals the number of conjugacy classes. Hence, the 
character table is a square $c \times c$ matrix. In order to decompose the 
$\rep{6}$ in \Eqref{eqn:decomposeintoPreps} we use 
$\chi_{\rep{6}}(g) = \chi_{\rep{a}}(g) + \chi_{\rep{b}}(g) + \ldots$ and the 
orthogonality of the rows of the character table (where the scalar product 
is defined over all elements of the conjugacy classes). In detail, for two 
irreducible representations $\rep{\alpha}$ and $\rep{\beta}$, we have
\begin{equation}
 \langle \rep{\alpha}, \rep{\beta}\rangle 
 ~=~ 
 \frac{1}{|P|} \sum_{g \in P} \chi_{\rep{\alpha}}(g) \overline{\chi_{\rep{\beta}}(g)} 
 ~=~ \left\{ \begin{array}{cl} 1 & \text{for }\rep{\alpha} = \rep{\beta}\;, \\ 
 0 & \text{for }\rep{\alpha} \neq \rep{\beta}\;, \end{array}\right.
\end{equation}
where the overline indicates complex conjugation and $|P|$ is the order of $P$. 
So for each conjugacy class $\left[g\right]$ we compute the character 
$\chi_{\rep{\xi}}(g)$ of the six--dimensional representation $\rep{6}$, now 
denoted by $\rep{\xi}$, and determine the multiplicities $n_i$ of the 
irreducible representation $\rep{\rho}_i$ in the decomposition,
\begin{equation}
\label{eqn:decompositionxi}
\rep{\xi} ~\rightarrow~ \bigoplus_{i=1}^c n_i\, \rep{\rho}_i  \quad\text{with}\quad n_i ~=~ \frac{1}{|P|} \sum_{g \in P} \chi_{\rep{\rho}_i}(g) \overline{\chi_{\rep{\xi}}(g)}\;.
\end{equation}
If $P$ is a subgroup of $\SU{3}$ this decomposition has to be of the kind 
\begin{equation}
\label{eqn:decompositionPfromSU3}
\rep{6} ~\rightarrow~ \rep{a} \oplus \crep{a}\;,
\end{equation}
where $\rep{a}$ denotes some (in general reducible) representation of $P$ 
originating from the $\rep{3}$ of $\SU{3}$ and $\crep{a}$ its complex conjugate 
(from $\crep{3}$ of $\SU{3}$). So, the first check is to see whether the 
decomposition (\ref{eqn:decompositionxi}) is of the form 
(\ref{eqn:decompositionPfromSU3}). Then we know at least 
$P \subset \text{U}(3)$. If this is possible, then there are in 
general many combinations to arrange the representations $\rep{\rho}_i$ of the 
decomposition (\ref{eqn:decompositionxi}) into a three--dimensional 
representation plus its complex conjugate. But in order to see that $P$ is a 
subgroup of $\text{(S)U}(3)$ it is necessary to find only one combination. 
However, one needs to know the explicit matrix representation of $\rep{a}$ in 
order to check that the determinant is $+1$. Then $P \subset \SU{3}$ and at 
least $\mathcal{N}=1$ SUSY survives the compactification of the heterotic 
string on the corresponding orbifold.

Let us make a short remark. If a point group is Abelian its generators can be 
diagonalized simultaneously. In this case, it is convenient to write them as 
so--called twist vectors $v=(v_1,v_2,v_3)$, three--dimensional vectors 
containing the three rotational angles $v_i$ in units of $2\pi$ in the three 
complex planes $i=1,2,3$. In this case, the check $P \subset \SU{3}$ is 
particular easy: $v_1+v_2+v_3={0 \mod 1}$ so that the determinant is $+1$. 
More precisely, it is always possible to choose the signs of the $v_i$ such
that they add to 0.
For example, the generator of the $\Z{7}$ point group corresponds to the twist 
vector $\frac{1}{7}(1,2,-3)$ with $\frac{1}{7}(1+2-3)=0$ such that 
$\Z{7} \subset\SU{3}$.

We use the software GAP \cite{GAP4} and the GAP package Repsn 
\cite{Dabbaghian:2004} for these computations. In detail, first we use GAP to 
uniquely identify the discrete group $P$ by the GAPID $\left[N,M\right]$, where 
$N$ denotes the order of the group and $M$ consecutively enumerates the 
discrete groups of order $N$. Then we perform the decomposition of the 
six--dimensional representation according to \Eqref{eqn:decompositionxi}. If 
the decomposition cannot be arranged according to 
\Eqref{eqn:decompositionPfromSU3} we know that $P$ is not a subgroup of 
$\SU{3}$. Otherwise, we create all combinations that fit with 
\Eqref{eqn:decompositionPfromSU3} and compute the explicit matrix 
representation using the GAP package Repsn.\footnote{In our case, Repsn 
automatically created unitary representations except for one case (point group 
$\PSL{3,2}$). In this case we had to transform the representation 
obtained by Repsn to a unitary one by hand.} Then we can easily compute the 
determinant of the generators of $P$ in the (reducible) representation $\rep{a}$.

\subsubsection*{Example: $\boldsymbol{S_3}$ point group}

As an example we consider $P=S_3$ and follow the steps in order to check that 
$S_3 \subset \SU{3}$. The $2262^\mathrm{nd}$ \QQ--class obtained from \Carat\ is 
generated by two $\GL{6,\ZZ}$ matrices, both of determinant $+1$,
\begin{equation}
\label{eqn:generatorsS3}
\vartheta_\basis{e}^{(\rep{\xi})} ~=~ \left(
\begin{array}{cccccc}
 1 &  1 &  0 & 0 &  0 &  0 \\
 0 & -1 &  0 & 0 &  0 &  0 \\
 0 &  0 & -1 & 0 &  0 &  0 \\
 0 &  0 & -1 & 1 &  0 &  0 \\
 0 &  0 &  0 & 0 & -1 &  0 \\
 0 &  0 &  0 & 0 &  0 & -1 \\
\end{array}
\right)\quad\text{and}\quad
\omega_\basis{e}^{(\rep{\xi})} ~=~ \left(
\begin{array}{cccccc}
  0 & 1 & 0 & 0 & 0 & 0 \\
 -1 &-1 & 0 & 0 & 0 & 0 \\
  0 & 0 &-1 & 1 & 0 & 0 \\
  0 & 0 &-1 & 0 & 0 & 0 \\
  0 & 0 & 0 & 0 & 1 & 0 \\
  0 & 0 & 0 & 0 & 0 & 1 \\
\end{array}
\right)\;.
\end{equation}
The group generated by these (non--commuting) matrices is identified by GAP as 
GAPID $\left[ 6, 1 \right]$ being $S_3$. $\vartheta_\basis{e}^{(\rep{\xi})}$ 
and $\omega_\basis{e}^{(\rep{\xi})}$ generate the six--dimensional (reducible) 
representation $\rep{\xi}$ of $S_3$. In what follows, we figure out how
this decomposes  into irreducible representations of $S_3$.

The character table of $S_3$ reads (in the ordering given by GAP)
\begin{equation}
\label{eqn:charactertableS3}
\begin{array}{c|ccc}
\text{irrep} & \left[\Id\right] & \left[\vartheta_\basis{e} \right] & \left[\omega_\basis{e}\right] \\
\hline
\rep{\rho}_1 &  1 &  1 &  1 \\
\rep{\rho}_2 &  1 & -1 &  1 \\
\rep{\rho}_3 &  2 &  0 & -1 \\
\end{array}
\end{equation}
where $\rep{\rho}_1$ denotes the singlet and $\rep{\rho}_2$ and $\rep{\rho}_3$ 
are a one-- and a two--dimensional (non--trivial) representation of $S_3$, 
respectively. Note that the conjugacy class $\left[\vartheta_\basis{e} \right]$ 
contains three elements while $\left[\omega_\basis{e}\right]$ contains two.
Furthermore, the characters of the six--dimensional representation $\rep{\xi}$ 
generated by \Eqref{eqn:generatorsS3} read
\begin{equation}
\chi_{\rep{\xi}}
~=~\left(\tr\mathbbm{1}_6,\tr\vartheta_\basis{e}^{(\rep{\xi})},\tr\omega_\basis{e}^{(\rep{\xi})}\right)
~=~ \left(6,-2,0\right).
\end{equation}
Comparing this to the character table in \Eqref{eqn:charactertableS3} we
find that  $\rep{\xi}$ decomposes into irreducible representations of $S_3$ as
\begin{equation}
\rep{\xi} ~\rightarrow~ 2 \rep{\rho}_2 \oplus 2 \rep{\rho}_3\;. 
\end{equation}
The only combination that fits into a three--dimensional representation is 
$\rep{\rho}_2 \oplus \rep{\rho}_3$. Using the GAP package Repsn we create the 
explicit matrix representation of this, resulting in
\begin{equation}
\vartheta^{(\rep{3})} ~=~ \left(
\begin{array}{ccc}
 -1 & 0 & 0 \\
  0 & 0 & 1 \\
  0 & 1 & 0 \\
\end{array}
\right)\quad\text{and}\quad
\omega^{(\rep{3})} ~=~ \left(
\begin{array}{ccc}
 1 & 0 & 0 \\
 0 & \exp\left(-\frac{2\pi\,\I}{3}\right) & 0 \\
 0 & 0 & \exp\left(\frac{2\pi\,\I}{3}\right) \\
\end{array}
\right)\;.
\end{equation}
As both generators have determinant $+1$, we see that $S_3 \subset\SU{3}$. 
Furthermore, since $\rep{3} \rightarrow \rep{\rho}_2 \oplus \rep{\rho}_3$ does 
not contain the trivial singlet $\rep{\rho}_1$, we see that $\mathcal{N}=1$ 
SUSY (and not more) is preserved by an $S_3$ orbifold compactification.

Recently, an explicit example of a non--Abelian orbifold based on $S_3$ has
been constructed \cite{Konopka:2011dt}. Among other things, such settings
feature, unlike Abelian orbifolds, rank reduction of the gauge symmetry already
at the string level.

\section{Results: classification of toroidal orbifolds}
\label{sec:orbclass}

We perform a systematic classification of space groups that keep (at least) 
$\mathcal{N}=1$ SUSY in four dimensions unbroken. As discussed in
\Secref{sec:class},  the amount of unbroken supersymmetry depends only on the
\QQ--class (i.e.\ point  group). Using \Carat\ we know that there are 7103
\QQ--classes in six  dimensions. Out of those, we find 60 \QQ--classes with
$\mathcal{N} \geq 1$  SUSY where 52 lead to precisely $\mathcal{N} = 1$, see 
\Tabref{tab:SummaryPointGroupsWithSUSY} for a summary of the results. The 60 
cases split into 22 Abelian and 38 non--Abelian \QQ--classes, where the Abelian 
cases were already known in the literature. By contrast, most of the 38 
non--Abelian \QQ--classes have not been used in orbifold compactifications
before. Starting from these 60  \QQ--classes we construct all possible \ZZ-- and
affine classes (i.e.\  lattices and roto--translations). In the following we
discuss  them in detail: Sections \ref{sec:AbelianClassification} and 
\ref{sec:NonAbelianClassification} are devoted to the Abelian and non--Abelian 
case, respectively.

\begin{table}[!ht]
\centering
\begin{tabular}{|cc|c|c|}
\hline
$\#$ of generators & $\#$ of SUSY  & Abelian & non--Abelian \\
\hline
\hline
 1               & $\mathcal{N}=4$ &       1 &  0 \\
                 & $\mathcal{N}=2$ &       4 &  0 \\
                 & $\mathcal{N}=1$ &       9 &  0 \\
\cline{3-4}
                 &                 &      14 &  0 \\
\hline
\hline
 2               & $\mathcal{N}=4$ &       0 &  0 \\
                 & $\mathcal{N}=2$ &       0 &  3 \\
                 & $\mathcal{N}=1$ &       8 & 32 \\
\cline{3-4}
                 &                 &       8 & 35 \\
\hline
\hline
 3               & $\mathcal{N}=4$ &       0 &  0 \\
                 & $\mathcal{N}=2$ &       0 &  0 \\
                 & $\mathcal{N}=1$ &       0 &  3 \\
\cline{3-4}
                 &                 &       0 &  3 \\
\hline
\hline
total:           & $\mathcal{N}=4$ &       1 &  0 \\
                 & $\mathcal{N}=2$ &       4 &  3 \\
                 & $\mathcal{N}=1$ &      17 & 35 \\
\cline{3-4}
                 &                 &      22 & 38 \\
\hline
\end{tabular}
\caption{Summary of the classification of all point groups with at least $\mathcal{N}=1$ SUSY. Out of 7103 cases obtained from \Carat\ there are 60 point groups with $\mathcal{N}\geq 1$ SUSY where 52 have exactly $\mathcal{N}=1$.}
\label{tab:SummaryPointGroupsWithSUSY}
\end{table}

\subsection{Abelian toroidal orbifolds}
\label{sec:AbelianClassification}

\subsubsection{Our results}

Restricting ourselves to Abelian point groups, we find 17 point groups with 
$\mathcal{N} = 1$ SUSY, four cases with $\mathcal{N} = 2$ and one case  (i.e.\
the trivial point group) with $\mathcal{N} = 4$ supersymmetry.  Next, we
classify all \ZZ-- and affine classes. For the 17 point groups with 
$\mathcal{N} = 1$ it turns out that there are in total 138 inequivalent space 
groups with Abelian point group and $\mathcal{N} = 1$. Many of them were
unknown  before. The results are summarized in 
\Tabref{tab:SummaryAbelianPointGroupsWithSUSY}. More details including  the
generators of the orbifolding group $G$, the nature of gauge symmetry  breaking
(i.e.\ local or non--local) and the Hodge numbers  $(h^{(1,1)},h^{(2,1)})$ can
be found in the Appendix in \Tabref{tab:complist}.  Furthermore, we have plotted
the 138 pairs of Hodge numbers in  \Figref{fig:hodgenumbers} in the Appendix,
visualizing the fact that  $h^{(1,1)} -h^{(2,1)}$ is always divisible by 6,
except for the case  $(h^{(1,1)},h^{(2,1)})=(20,0)$. Note that this does not say
that Standard Models with three generations of quarks and leptons are
impossible, due to the possibility of introducing so--called discrete Wilson lines
\cite{Dixon:1986jc,Ibanez:1986tp} and/or discrete torsion
\cite{Vafa:1986wx,Font:1988mk,Vafa:1994rv,Sharpe:2000ki,Gaberdiel:2004vx,Ploger:2007iq}.

At this point, a comment on a statement in DW~\cite{Donagi:2008xy} appears
appropriate. The models obtained in the free
fermionic construction (such as \cite{Cleaver:1998sa}) are claimed to be
related to $\ZxZ{2}{2}$ orbifolds. DW~\cite{Donagi:2008xy} conclude from the fact
that their classification does not exhibit settings with $h^{(1,1)} -h^{(2,1)}$
equal to three, that the free fermionic models, hence, cannot have a geometric
interpretation. However, as pointed out in \cite{Ibanez:1986tp} and also in
\cite{Ploger:2007iq}, discrete Wilson lines and/or (generalized) discrete
torsion allows us to control the number of generations. We do not know whether
some of the boundary conditions in the free fermionic construction correspond to
such backgrounds. On the other hand, the existing three generation models based
on $\ZxZ{2}{2}$ orbifolds \cite{Forste:2004ie,Blaszczyk:2009in,Kappl:2010yu}
make use of discrete Wilson lines and have, at the same time, a geometric
interpretation. This might mean that the models in the free fermionic
constructions may also be `geometric'.

\begin{table}[ht]
\centering
\begin{tabular}{|c||c||c|c|c||c|c|}
\hline
label of        & twist                                         & GAPID       & \Carat      & \Carat  & $\#$ of          & \!$\#$ of affine\! \\
\QQ--class      & vector(s)                                     &             & symbol     & index  & \!\ZZ--classes\! & classes\\
\hline
\hline
$\Z{3}$         & $\frac{1}{3}(1,1,-2)$                         &   \GAP{3,1} &    min.290 & $1965$ &  $1$ &  $1$ \\
$\Z{4}$         & $\frac{1}{4}(1,1,-2)$                         &   \GAP{4,1} &    min.201 & $4667$ &  $3$ &  $3$ \\
$\Z{6}$--I       & $\frac{1}{6}(1,1,-2)$                         &   \GAP{6,2} &    min.296 & $1997$ &  $2$ &  $2$ \\
$\Z{6}$--II      & $\frac{1}{6}(1,2,-3)$                         &   \GAP{6,2} &    min.403 &  $944$ &  $4$ &  $4$ \\
$\Z{7}$         & $\frac{1}{7}(1,2,-3)$                         &   \GAP{7,1} &    min.665 & $2950$ &  $1$ &  $1$ \\
$\Z{8}$--I       & $\frac{1}{8}(1,2,-3)$                         &   \GAP{8,1} &    min.475 & $5600$ &  $3$ &  $3$ \\
$\Z{8}$--II      & $\frac{1}{8}(1,3,-4)$                         &   \GAP{8,1} &    min.467 & $5567$ &  $2$ &  $2$ \\
$\Z{12}$--I      & $\frac{1}{12}(1,4,-5)$                        &  \GAP{12,2} &    min.562 & $3346$ &  $2$ &  $2$ \\
$\Z{12}$--II     & $\frac{1}{12}(1,5,-6)$                        &  \GAP{12,2} &    min.553 & $3307$ &  $1$ &  $1$ \\
$\ZxZ{2}{2}$    & $\frac{1}{2}(0,1,-1)\;,\;\frac{1}{2}(1,0,-1)$ &   \GAP{4,2} &    min.185 & $4625$ & $12$ & $35$ \\
$\ZxZ{2}{4}$    & $\frac{1}{2}(0,1,-1)\;,\;\frac{1}{4}(1,0,-1)$ &   \GAP{8,2} &    min.258 & $2377$ & $10$ & $41$ \\
$\ZxZ{2}{6}$--I  & $\frac{1}{2}(0,1,-1)\;,\;\frac{1}{6}(1,0,-1)$ &  \GAP{12,5} & group.2702 &  $871$ &  $2$ &  $4$ \\
$\ZxZ{2}{6}$--II & $\frac{1}{2}(0,1,-1)\;,\;\frac{1}{6}(1,1,-2)$ &  \GAP{12,5} &    min.424 & $1745$ &  $4$ &  $4$ \\
$\ZxZ{3}{3}$    & $\frac{1}{3}(0,1,-1)\;,\;\frac{1}{3}(1,0,-1)$ &   \GAP{9,2} &    min.429 & $1964$ &  $5$ & $15$ \\
$\ZxZ{3}{6}$    & $\frac{1}{3}(0,1,-1)\;,\;\frac{1}{6}(1,0,-1)$ &  \GAP{18,5} & group.3567 & $1759$ &  $2$ &  $4$ \\
$\ZxZ{4}{4}$    & $\frac{1}{4}(0,1,-1)\;,\;\frac{1}{4}(1,0,-1)$ &  \GAP{16,2} &    min.278 & $2629$ &  $5$ & $15$ \\
$\ZxZ{6}{6}$    & $\frac{1}{6}(0,1,-1)\;,\;\frac{1}{6}(1,0,-1)$ & \GAP{36,14} & group.3664 & $1859$ &  $1$ &  $1$ \\
\hline
\hline
\multicolumn{5}{|l||}{\# of Abelian $\mathcal{N}=1$}                                                  & $60$ & $138$ \\
\hline
\end{tabular}
\caption{Summary of all space groups with Abelian point group and 
$\mathcal{N}=1$ SUSY. Columns \# 3, 4 and 5 identify the \QQ--classes: ``GAPID'' 
is obtained using the command \texttt{IdGroup} in GAP, ``\Carat\ symbol'' using 
the \Carat\ command \texttt{Q\_catalog} and, finally, ``\Carat\ index'' gives the 
index in the list of all 7103 \QQ--classes obtained from \Carat.} 
\label{tab:SummaryAbelianPointGroupsWithSUSY}
\end{table}

The results are also available as input for the {\small\tt orbifolder} 
\cite{Nilles:2011aj}, a tool to study the low energy phenomenology of heterotic 
orbifolds. We have created input files for the {\small\tt orbifolder}, which we 
have made available at
\begin{center}
\url{http://einrichtungen.physik.tu-muenchen.de/T30e/codes/ClassificationOrbifolds/}\;.
\end{center}
There is a geometry file for each of  the 138 affine classes, and one model 
file per \QQ--class, that contains a  model with standard embedding for each of 
the corresponding affine classes in  that \QQ--class.

In addition, we find 23 inequivalent space groups (i.e.\ affine classes) 
with Abelian point group and $\mathcal{N} = 2$. These space groups are based on 
the well--known four Abelian point groups $\Z{2}$, $\Z{3}$, $\Z{4}$ and 
$\Z{6}$. However, the inequivalent lattices and roto--translations were unknown 
before. They are summarized in \Tabref{tab:SummaryNLargerOne}.

\begin{table}[ht]
\centering
\begin{tabular}{|c||c|c|c||c|c|}
\hline
label of                                    & GAPID        & \Carat      & \Carat  & $\#$ of      & $\#$ of affine \\
\QQ--class                                  &              & symbol     & index  & \ZZ--classes & classes\\
\hline
\hline
$\Z{2}$                                     &    \GAP{2,1} &    min.174 &    $5$ &  $3$ &  $5$ \\
$\Z{3}$                                     &    \GAP{3,1} &    min.291 & $1968$ &  $3$ &  $5$ \\
$\Z{4}$                                     &    \GAP{4,1} &    min.202 & $4668$ &  $3$ &  $9$ \\
$\Z{6}$                                     &    \GAP{6,2} & group.1611 & $1970$ &  $1$ &  $4$ \\
\hline
\hline
\multicolumn{4}{|l||}{\# of Abelian $\mathcal{N}=2$}                               & $10$ & $23$ \\
\hline
\end{tabular}
\caption{Summary of all space groups with $\mathcal{N} > 1$ SUSY for Abelian 
point groups $P$. In addition, there is the trivial \QQ--class with 
$\mathcal{N}=4$ SUSY (i.e.\ GAPID \GAP{1,1}, \Carat\ symbol min.170, \Carat\ 
index $2709$) with one \ZZ-- and one affine class.}
\label{tab:SummaryNLargerOne}
\end{table}

\subsubsection{Previous classifications}

There are several attempts in the literature to classify six--dimensional 
$\mathcal{N}=1$ SUSY preserving Abelian toroidal orbifolds. For example, 
Bailin and Love \cite{Bailin:1999nk} give a classification for $\Z{N}$ 
orbifolds using root lattices of semi--simple Lie algebras of rank six as 
lattices $\Lambda$ and the (generalized) Coxeter element as the generator of 
the point group $P$. However, as also discussed in \Apref{app:bravaislie}, they 
overcount the geometries and, in addition, miss a few cases. A detailed 
comparison to our results can be found in \Tabref{tab:ZNComparison}.

For $\ZxZ{2}{2}$ orbifolds there have been two approaches for the
classification  of geometries. In the first one, the classification is  based on
Lie lattices \cite{Forste:2006wq}, see also \cite{Kimura:2007ey}. Again,  this
classification is somewhat incomplete: it misses four lattices and, in 
addition, neglects the possibility of roto--translations. In a second approach 
by DW~\cite{Donagi:2008xy} (based on \cite{Donagi:2004ht}), a 
classification for $\ZxZ{2}{2}$ is given, which, as we find, is complete (but 
overcounts one case), see \Tabref{tab:z2z2} for a comparison.

Furthermore, based on the strategy of DW~\cite{Donagi:2008xy}, 
there is an (incomplete) classification of $\ZxZ{N}{N}$ for  $N=3,\,4\text{ and
}6$ \cite{Dillies:2006yb}. For both $\ZxZ{3}{3}$ and  $\ZxZ{4}{4}$ they identify
8 out of 15 affine classes (compare Section 2.3 of  \cite{Dillies:2006yb} to our
\Tabref{tab:complist}). Their Hodge numbers  agree with our findings except for
their case IV.7 (i.e.\ $\ZxZ{4}{4}$ with  $(38,0)$). Finally, in the case of
$\ZxZ{6}{6}$ they correctly identify that  there is only one possible geometry
but their Hodge numbers disagree with ours, i.e.\  they find $(80,0)$ and
we have $(84,0)$.

\begin{table}[!ht]
\begin{center}
\begin{tabular}{|l|c|l|}
\hline
\multicolumn{1}{|c|}{\QQ--class} &
\multicolumn{1}{|c|}{\ZZ--class} &
\multicolumn{1}{|c|}{corresponding root lattice(s)}\\
\hline
\hline
$\Z{3}$              & 1 & $\SU{3}^3$ $\phantom{I^{I^{[I]}}}$\\
\hline
$\Z{4}$              & 1 & $\SO{5}^2 \times \SU{2}^2$ $\phantom{I^{I^{[I]}}}$\\
\cline{2-3}
                     & 2 & $\SO{5} \times \SU{4} \times \SU{2}$\\
\cline{2-3}
                     & 3 & $\SU{4}^2$ $\phantom{I^{I^{[I]}}}$\\
\hline
$\Z{6}$--I   & 1 & $\left(\mathrm{G}_2\right)^2 \times \SU{3}$\;\text{and}\;$\left(\SU{3}^{[2]}\right)^2 \times \SU{3}$ $\phantom{I^{I^{[I]}}}$\\
\cline{2-3}
                     & 2 &  --- \\
\hline
$\Z{6}$--II  & 1 & $\mathrm{G}_2 \times \SU{3} \times \SU{2}^2$ \;\text{and}\; $\SU{3}^{[2]} \times \SU{3} \times \SU{2}^2$ $\phantom{I^{I^{[I]}}}$\\
\cline{2-3}
                     & 2 & --- \\
\cline{2-3}
                     & 3 & $\SO{8} \times \SU{3}$ \;\text{and}\; $\SO{7} \times \SU{3} \times \SU{2}$ \;\text{and} \\
                     &   & $\SU{4}^{[2]} \times \SU{3} \times \SU{2}$\\
\cline{2-3}
                     & 4 & $\SU{6} \times \SU{2}$\\
\hline
$\Z{7}$              & 1 & $\SU{7}$\\
\hline
$\Z{8}$--I   & 1 & $\SO{9} \times \SO{5}$\;\text{and}\; $\SO{8}^{[2]} \times \SO{5}$ $\phantom{I^{I^{[I]}}}$\\
\cline{2-3}
                     & 2 & --- \\
\cline{2-3}
                     & 3 & --- \\
\hline
$\Z{8}$--II  & 1 & $\SO{8}^{[2]} \times \SU{2}^2$ \;\text{and}\; $\SO{9} \times \SU{2}^2$ $\phantom{I^{I^{[I]}}}$\\
\cline{2-3}
                     & 2 & $\SO{10} \times \SU{2}$\\
\hline
$\Z{12}$--I  & 1 & $\mathrm{F}_4 \times \SU{3}$ \;\text{and}\; $\SO{8}^{[3]} \times \SU{3}$ $\phantom{I^{I^{[I]}}}$\\
\cline{2-3}
                     & 2 & $\mathrm{E}_6$\\
\hline
$\Z{12}$--II & 1 & $\SO{4} \times \mathrm{F}_4$ \;\text{and}\; $\SO{8}^{[3]} \times \SU{2}^2$ $\phantom{I^{I^{[I]}}}$\\
\hline
\end{tabular}
\caption{Matching between our classification of $\Z{N}$ space groups and the 
traditional notation of lattices as root lattices of semi--simple Lie algebras 
of rank six, see e.g.\ Table 3 of \cite{Bailin:1999nk} and Table
D.1 of \cite{RamosSanchez:2008tn}. Cases previously not known are indicated 
with a dash.}
\label{tab:ZNComparison}
\end{center}
\end{table}

\begin{table}[ht]
\centering
\begin{tabular}{|c|c|c|c|c|c|c|c|c|}
\hline
Here & Donagi                      & F{\"o}rste et al.    & $\pi_1$ && Here & Donagi                      & F{\"o}rste et al.    & $\pi_1$\\ 
     & et al. \cite{Donagi:2008xy} & \cite{Forste:2006wq} &         &&      & et al. \cite{Donagi:2008xy} & \cite{Forste:2006wq} &        \\
\hline
\hline
1--1   &  0--1  & $\SU{2}^6$                         & 0 && 5--3   &  1--2&---&0\\
1--2   &  0--2  & ---                                & 0 && 5--4   &  1--4&---&A\\
1--3   &  0--3  & ---                                & A && 5--5   &  1--5&---&S\\
\cline{6-9}
1--4   &  0--4  & ---                                & S && 6--1   &  2--6& $\SU{3}^2\times\SU{2}^2$--II&0\\
\cline{1-4}
2--1   &  1--6  & $\SU{3}\times\SU{2}^4$             & 0 && 6--2   &  2--7&---&C\\
2--2   &  1--8  & ---                                & 0 && 6--3   &  2--8&---&A\\
\cline{6-9}
2--3   &  1--10 & ---                                & A && 7--1   &  3--3&---&0\\
2--4   &  1--7  & ---                                & C && 7--2   &  3--4&---&C\\
\cline{6-9}
2--5   &  1--9  & ---                                & A && 8--1   &  4--1&---&0\\
\cline{6-9}
2--6   &  1--11 & ---                                & S && 9--1   &  2--3& $\SU{4}\times\SU{3}\times\SU{2}$&C\\
\cline{1-4}
3--1   &  2--9  & ---                                & 0 && 9--2   &  2--5&---&D\\
3--2   &  2--10 & ---                                & 0 && 9--3   &  2--4&---&0\\
\cline{6-9}
3--3   &  2--11 & ---                                & A && 10--1  &  3--5&---&C\\
3--4   &  2--12 & ---                                & S && 10--2  &  3--6&---&0\\
\cline{1-4}\cline{6-9}
4--1   &  2--13 & $\SU{3}^2\times\SU{2}^2$--I        & 0 && 11--1  &  3--1$\equiv$3--2& $\SU{3}^3$&0\\
\cline{6-9}
4--2   &  2--14 & ---                                & D && 12--1  &  2--1& $\SU{4}^2$&D\\
\cline{1-4}
5--1   &  1--1  & $\SU{4}\times\SU{2}^3$             & C && 12--2  &  2--2 &---&C\\
5--2   &  1--3  & ---                                & C &&        &       &   & \\
\hline
\end{tabular}
\caption{Comparison of the affine classes of $\ZxZ{2}{2}$ between our 
classification and the ones in \cite{Donagi:2008xy} and \cite{Forste:2006wq}. 
In our case, the two numbers enumerate the \ZZ-- and affine classes, respectively.}
\label{tab:z2z2}
\end{table}

\subsubsection{Fundamental groups}

The fundamental group of a toroidal orbifold with space group $S$ is 
given as \cite{Dixon:1986jc,Brown:2002}
\begin{equation}\label{eq:fg}
\pi_1~=~S/\langle F\rangle\;,
\end{equation}
where $\langle F\rangle$ is the group generated by those space group elements 
that leave some points fixed.

The fundamental groups of most of the Abelian orbifolds discussed here are 
trivial, for in those cases $\langle F \rangle\equiv S$. The only non--trivial 
cases are the following (see \Tabref{tab:complist} in the Appendix):
\begin{itemize}
\item 21 space groups from the $\ZxZ{2}{2}$ \QQ--class as already calculated 
      in \cite{Donagi:2008xy}. See \Tabref{tab:z2z2}, where
  \begin{itemize}
  \item 0 means a trivial fundamental group
  \item $S$ means the fundamental group equals the space group (no 
        fixed points, hence $\langle F\rangle=\{\Id\}$)
  \item $A$ means a $\ZZ_2\ltimes\ZZ^2$ fundamental group
  \item $C$ means a $\ZZ_2$ fundamental group
  \item $D$ means a $(\ZZ_2)^2$ fundamental group
  \end{itemize}

\item 6 space groups from the $\ZxZ{2}{4}$ \QQ--class. In detail, the 
      affine classes 1--6, 2--4, 3--6, 4--4, 6--5 and 8--3 posses a $\Z{2}$ 
      fundamental group.
\item 4 space groups from the $\ZxZ{3}{3}$ \QQ--class. In detail, the 
      affine classes 1--4, 2--4, 3--3 and 4--3 posses a $\Z{3}$ fundamental 
      group.
\end{itemize}

Elements of the space group that leave no fixed points are called freely
acting. A non--trivial fundamental group signals the presence of 
non--decomposable freely acting elements in the space group, i.e.\ freely
acting elements that cannot be written as a combination of non--freely
acting elements. In the cases $\ZxZ{2}{4}$ and $\ZxZ{3}{3}$, the non--decomposable
freely acting elements belong to the orbifolding group. On the other hand,
for $\ZxZ{2}{2}$ those elements are pure lattice translations in the
cases $C$ and $D$, while in the cases $A$ they are both pure lattice 
translations and elements of the orbifolding group.

In the context of heterotic compactifications, the phenomenologically appealing 
feature of non--local GUT symmetry breaking is due to the presence of 
non--decomposable freely acting space group elements with a non--trivial gauge 
embedding. In total we find 31 affine classes based on Abelian point groups 
with non--trivial fundamental groups. These cases are of special interest, and 
their phenomenology will be studied elsewhere.

\subsection{\texorpdfstring{Non--Abelian toroidal orbifolds}{Non-Abelian toroidal orbifolds}}
\label{sec:NonAbelianClassification}

Orbifolds with non--Abelian point groups have not been studied systematically 
up to now and the literature is limited to examples only. For example, in the 
context of free fermionic constructions compact models based on $S_3$, $D_4$ 
and $A_4$ point groups have been constructed \cite{Kakushadze:1996hj}. 
Furthermore, non--compact examples of the form $\mathbbm{C}^3/\Gamma$ with non--Abelian 
$\Gamma \subset \SU{3}$ focusing on $\Gamma = \Delta(3n^2)$ or $\Delta(6n^2)$ 
have been discussed in \cite{Muto:1998na} and some related work has been
carried  out for IIB superstring theory on $\text{AdS}_5\times \mathbbm{S}^5/\Gamma$ with 
non--Abelian $\Gamma \subset \SU{3}$ of order up to 31 \cite{Frampton:2000mq}.

Our classification results in 35 point groups with $\mathcal{N} = 1$ SUSY and  
three cases with $\mathcal{N} = 2$ SUSY, see \Tabref{tab:NonAbelianPointGroups} in
\Apref{app:results}. Surprisingly, the order of non--Abelian point groups has a 
much wider range compared to the Abelian case. For example, the point group
$\Delta(216)$ has order 216.

Next, we classify all \ZZ-- and affine classes. It turns out that there are 
in total 331 inequivalent space groups with non--Abelian point group and 
$\mathcal{N} = 1$ SUSY and 27 inequivalent space groups with non--Abelian point 
group and $\mathcal{N} = 2$. Most of them were unknown before. The results are 
summarized in \Tabref{tab:SummaryNLargerOneNonAbelian} and 
\Tabref{tab:SummaryNOneNonAbelian}.

\begin{table}[ht]
\centering
\begin{tabular}{|c||c|c|c||c|c|}
\hline
label of                                    & GAPID        & \Carat      & \Carat  & $\#$ of      & $\#$ of affine \\
\QQ--class                                  &              & symbol     & index  & \ZZ--classes & classes\\
\hline
\hline
$Q_8$                                       &    \GAP{8,4} &    min.487 & $5750$ &  $5$ & $20$ \\
$\Dic{3}$                                   &   \GAP{12,1} &    min.565 & $3374$ &  $1$ &  $3$ \\
$\SL{2,3}$--II                              &   \GAP{24,3} & group.4493 & $5669$ &  $1$ &  $4$ \\
\hline
\hline
\multicolumn{4}{|l||}{\# of non--Abelian $\mathcal{N}=2$}                          &  $7$ & $27$ \\
\hline
\end{tabular}
\caption{Summary of all space groups with $\mathcal{N} > 1$ SUSY for 
non--Abelian $P$.}
\label{tab:SummaryNLargerOneNonAbelian}
\end{table}

\subsubsection*{Example: $\boldsymbol{D_6}$ Orbifold}

Let us consider the $\mathbbm{T}^6/D_6$ orbifold. $D_6$ is a non--Abelian finite 
group of order 12. The (reducible) three--dimensional representation is 
generated by
\begin{equation}
\vartheta^{(\rep{3})} ~=~ \left(
\begin{array}{ccc}
-1 & 0 & 0 \\
 0 & 0 & 1 \\
 0 & 1 & 0
\end{array}
\right)\quad\text{and}\quad
\omega^{(\rep{3})} ~=~ \left(
\begin{array}{ccc}
 1 & 0 & 0 \\
 0 & \mathrm{e}^{2\pi\,\I\,\frac{1}{6}} & 0 \\
 0 & 0 & \mathrm{e}^{-2\pi\,\I\,\frac{1}{6}}
\end{array}
\right)\;,
\end{equation}
and one can see that $D_6 \subset\SU{3}$. In terms of irreducible 
representations of $D_6$ it decomposes as $\rep{3} \rightarrow \rep{2} \oplus 
\rep{1'}$, where $\rep{1'}$ is a non--trivial, one--dimensional representation 
of $D_6$. Hence, we find $\mathcal{N} = 1$ SUSY in 4D.

There are two inequivalent lattices (i.e.\ two \ZZ--classes) and in total eight 
affine classes, see \Tabref{tab:SummaryNOneNonAbelian}. For example, consider 
the space group generated by $\left(\vartheta^{(\rep{3})}, 0\right)$, 
$\left(\omega^{(\rep{3})}, 0\right)$ and the lattice
\begin{eqnarray}
e_1 & = & \left(1,0,0\right)\;, \quad e_2 ~=~ \left(\I,0,0\right)\;, \\ 
e_3 & = & \left(0,1,0\right)\;, \quad e_4 ~=~ \left(0,\mathrm{e}^{2\pi\,\I\,\frac{1}{3}},0\right)\;, \\
e_5 & = & \left(0,0,1\right)\;, \quad e_6 ~=~ \left(0,0,\mathrm{e}^{2\pi\,\I\,\frac{1}{3}}\right)\;.
\end{eqnarray}
As $D_6$ has six conjugacy classes, the $\mathbbm{T}^6/D_6$ orbifold has 
$6-1=5$ twisted sectors, all of them have fixed planes and hence are 
$\mathcal{N} = 2$ subsectors.

\begin{longtable}{|c||c|c|c||c|c|}
\hline
label of                                    & GAPID        & \Carat      & \Carat  & $\#$ of      & $\#$ of affine \\
\QQ--class                                  &              & symbol     & index  & \ZZ--classes & classes\\
\hline
\hline
\endfirsthead
\hline
label of                                    & GAPID        & \Carat      & \Carat  & $\#$ of      & $\#$ of affine \\
\QQ--class                                  &              & symbol     & index  & \ZZ--classes & classes\\
\hline
\hline
\endhead
\hline
\multicolumn{6}{|r|}{continued \ldots}\\
\hline
\endfoot
\endlastfoot
\hline
\hline
$S_3$                                       &    \GAP{6,1} &    min.300 & $2262$ &  $6$ & $11$ \\
$D_4$                                       &    \GAP{8,3} &    min.207 & $4682$ &  $9$ & $48$ \\
$A_4$                                       &   \GAP{12,3} &    min.430 & $4893$ &  $9$ & $15$ \\
$D_6$                                       &   \GAP{12,4} & group.1637 & $2258$ &  $2$ &  $8$ \\
$\Z{8}\rtimes\Z{2}$                         &   \GAP{16,6} &    min.506 & $6222$ &  $6$ & $18$ \\
$QD_{16}$                                   &   \GAP{16,8} & group.4474 & $5650$ &  $4$ & $14$ \\
$(\Z{4}\times\Z{2})\rtimes\Z{2}$            &  \GAP{16,13} & group.4469 & $5645$ &  $5$ & $55$ \\
$\Z{3}\times S_3$                           &   \GAP{18,3} &    min.613 & $4235$ &  $6$ & $16$ \\
Frobenius $T_7$                             &   \GAP{21,1} &    min.664 & $2935$ &  $3$ &  $3$ \\
$\Z{3}\rtimes\Z{8}$                         &   \GAP{24,1} &    min.511 & $6266$ &  $1$ &  $1$ \\
$\SL{2,3}$--I                               &   \GAP{24,3} &    min.536 & $6743$ &  $4$ &  $7$ \\
$\Z{4}\times S_3$                           &   \GAP{24,5} & group.5943 & $3414$ &  $1$ &  $2$ \\
$\left(\Z{6}\times\Z{2}\right)\rtimes\Z{2}$ &   \GAP{24,8} & group.5937 & $3408$ &  $2$ &  $6$ \\
$\Z{3}\times D_4$                           &  \GAP{24,10} &    min.616 & $4326$ &  $2$ &  $2$ \\
$\Z{3}\times Q_8$                           &  \GAP{24,11} &    min.528 & $6735$ &  $2$ &  $2$ \\
$S_4$                                       &  \GAP{24,12} & group.3770 & $4895$ &  $6$ & $19$ \\
$\Delta(27)$                                &   \GAP{27,3} &    min.659 & $2864$ &  $3$ & $10$ \\
$\left(\Z{4}\times\Z{4}\right)\rtimes\Z{2}$ &  \GAP{32,11} & group.5125 & $6337$ &  $5$ & $30$ \\
$\Z{3}\times \left(\Z{3}\rtimes\Z{4}\right)$&   \GAP{36,6} &    min.620 & $4353$ &  $1$ &  $1$ \\
$\Z{3}\times A_4$                           &  \GAP{36,11} &    min.661 & $2875$ &  $3$ &  $3$ \\
$\Z{6}\times S_3$                           &  \GAP{36,12} & group.6834 & $4356$ &  $2$ &  $4$ \\
$\Delta(48)$                                &   \GAP{48,3} &    min.651 & $2774$ &  $4$ &  $8$ \\
$\GL{2,3}$                                  &  \GAP{48,29} & group.4532 & $5713$ &  $1$ &  $4$ \\
$\SL{2,3}\rtimes\Z{2}$                      &  \GAP{48,33} & group.4531 & $5712$ &  $1$ &  $3$ \\
$\Delta(54)$                                &   \GAP{54,8} & group.7587 & $2897$ &  $3$ & $10$ \\
$\Z{3}\times\SL{2,3}$                       &  \GAP{72,25} & group.5746 & $6988$ &  $1$ &  $2$ \\
$\Z{3}\times\text{GAPID}\left[24,8\right]$  &  \GAP{72,30} & group.7007 & $4533$ &  $1$ &  $1$ \\
$\Z{3}\times S_4$                           &  \GAP{72,42} & group.7614 & $2924$ &  $3$ &  $3$ \\
$\Delta(96)$                                &  \GAP{96,64} & group.7498 & $2802$ &  $4$ & $12$ \\
$\SL{2,3}\rtimes\Z{4}$                      &  \GAP{96,67} & group.5290 & $6512$ &  $1$ &  $2$ \\
$\Sigma(36\phi)$                            & \GAP{108,15} & group.7500 & $2806$ &  $2$ &  $4$ \\
$\Delta(108)$                               & \GAP{108,22} & group.7504 & $2810$ &  $1$ &  $1$ \\
$\PSL{3,2}$                                 & \GAP{168,42} & group.7622 & $2934$ &  $1$ &  $3$ \\
$\Sigma(72\phi)$                            & \GAP{216,88} & group.7540 & $2846$ &  $2$ &  $2$ \\
$\Delta(216)$                               & \GAP{216,95} & group.7545 & $2851$ &  $1$ &  $1$ \\
\hline
\hline
\multicolumn{4}{|l||}{\# of non--Abelian $\mathcal{N}=1$}                          &$108$ & $331$ \\
\hline
\caption{Summary of all space groups with non--Abelian point group and $\mathcal{N}=1$ SUSY.}
\label{tab:SummaryNOneNonAbelian}
\end{longtable}

\section{Summary and Discussion}
\label{sec:Discussion}

We have classified all symmetric orbifolds that give $\mathcal{N}\ge1$
supersymmetry in four dimensions. Our main results are as follows:
\begin{enumerate}
  \item In total we find $60$ \QQ--classes (point groups) that lead to 
$\mathcal{N}\ge1$ SUSY.
\item These \QQ--classes decompose in 
\begin{itemize} 
\item $22$ with an Abelian point group with one or two generators, i.e.\
$\Z{N}$ or $\ZxZ{N}{M}$, out of which $17$ lead to exactly
$\mathcal{N}=1$ SUSY, and
\item $38$ with a non--Abelian point group with two or three generators,
such as $S_3$ or $\Delta(216)$, out of which $35$ lead to 
exactly $\mathcal{N}=1$ SUSY.
\end{itemize}
\setcounter{dislist}{\value{enumi}}
\end{enumerate}
That is, there are 52 \QQ--classes that can lead to models yielding the
supersymmetric standard model.

As we have explained in detail, \QQ--classes (or point groups) can come
with inequivalent lattices, classified by the so--called \ZZ--classes. In the
traditional  orbifold literature, \ZZ--classes are given by Lie lattices and a
given  choice fixes an orbifold geometry. However, as we have  pointed out, not
all lattices can be described by Lie lattices.

Our results on \QQ--classes potentially relevant for supersymmetric model
building are as follows.
\begin{enumerate}\setcounter{enumi}{\value{dislist}}
\item We find that there are $186$ \ZZ--classes, or, in other words, orbifold
geometries that lead to $\mathcal{N}\ge1$ SUSY. 
\item These \ZZ--classes decompose in 
\begin{itemize} 
\item $71$ with an Abelian point group, out of which $60$ lead to exactly
$\mathcal{N}=1$ SUSY, and
\item $115$ with a non--Abelian point group, out of which $108$ lead to 
exactly $\mathcal{N}=1$ SUSY.
\end{itemize}
\setcounter{dislist}{\value{enumi}}
\end{enumerate}
Furthermore, space groups can be extended by so--called roto--translations, 
a combination of a twist and a (non--lattice) translation. We provide a full 
classification of all roto--translations in terms of affine classes, which are, 
as we discuss, the most suitable objects to classify inequivalent space groups.
\begin{enumerate}\setcounter{enumi}{\value{dislist}}
\item We find 520 affine classes that lead to $\mathcal{N}\ge1$ SUSY. 
\item These affine classes decompose in 
\begin{itemize} 
\item $162$ with an Abelian point group, out of which $138$ lead to exactly 
$\mathcal{N}=1$ SUSY, and
\item $358$ with a non--Abelian point group, out of which $331$ lead to 
exactly $\mathcal{N}=1$ SUSY.
\end{itemize}
\end{enumerate}

An important aspect of our classification is that we provide the data 
for all 138 space groups with Abelian point group and $\mathcal{N}=1$ SUSY 
required to construct the corresponding models with the C++ orbifolder
\cite{Nilles:2011aj}. Among other things, this allows one to obtain a
statistical survey of the properties of the models, which has so far only been
performed for the $\Z{6}$--II orbifold \cite{Lebedev:2008un}.

Our classification also has conceivable importance for phenomenology. For
instance, one of the questions is how the ten--dimensional gauge group 
(i.e.\ $\E8\times\E8$ or $\SO{32}$) of the heterotic string gets broken by 
orbifolding. In most of the models discussed so far, the larger symmetry gets 
broken locally at some fixed point. Yet it has been argued that `non--local' 
GUT symmetry breaking, as utilized in the context of smooth compactifications 
of the heterotic string 
\cite{Bouchard:2005ag,Braun:2005ux,Braun:2011ni,Anderson:2012yf}, has certain 
phenomenological advantages \cite{Hebecker:2004ce,Anandakrishnan:2011zn}. 
Explicit MSSM candidate models, based on the DW classification, featuring 
non--local GUT breaking have been constructed recently 
\cite{Blaszczyk:2009in,Kappl:2010yu}. As we have seen, there are $31$ 
affine classes of space groups, based on the \QQ--classes $\ZxZ{2}{2}$,
$\ZxZ{2}{4}$ and $\ZxZ{3}{3}$, that lead to an orbifold with a non--trivial
fundamental group, thus allowing us to introduce a Wilson line that breaks the
GUT symmetry. In other words, we have identified a large set of geometries that
can give rise to non--local GUT breaking. This might also allow for a
dynamical stabilization of some of the moduli in the early universe, similar
as in toroidal compactifications \cite{Brandenberger:1988aj}.

In this study, we have focused on symmetric toroidal orbifolds, which have
a rather clear geometric interpretation, such that crystallographic methods can
be applied in a straightforward way. We have focused on the geometrical aspects.
On the other hand, it is known that background fields, i.e.\ the so--called
discrete Wilson lines \cite{Ibanez:1986tp} and discrete torsion
\cite{Vafa:1986wx,Vafa:1994rv,Sharpe:2000ki,Gaberdiel:2004vx,Ploger:2007iq},
play a crucial role in model building. It will be interesting to work out the
conditions on such background fields in the geometries of our classification.
Further, it is, of course, clear that there are other orbifolds, such as
asymmetric and/or non--toroidal orbifolds, whose classification is beyond the
scope of this study. Let us also mention, we implicitly assumed that the radii
are away from the self--dual point. It might be interesting to study what
happens if one sends one or more $T$--moduli to the self--dual values. In this
case one may make contact with the free fermionic formulation, where also
interesting models have been constructed \cite{Cleaver:1998sa}.
In addition, our results may also be applied to compactifications of type II 
string theory on orientifolds (see e.g.\ 
\cite{Gmeiner:2005vz,Douglas:2006xy,Gmeiner:2008xq} for some interesting 
models and \cite{Blumenhagen:2006ci} for a review).

\subsection*{Acknowledgments}

We would like to thank Pascal Vaudrevange for useful discussions. 
M.F.\ would like to thank Wilhelm Plesken for helpful advice regarding the use 
of \Carat. J.T.\ would like to thank Sebastian Konopka for very useful
discussions.  This work was partially supported by the Deutsche 
Forschungsgemeinschaft (DFG) through the cluster of excellence ``Origin and 
Structure of the Universe'' and the Graduiertenkolleg ``Particle Physics at the
Energy Frontier of New Phenomena''. P.V.\ is supported by SFB grant 676.  This
research was done in the context of the ERC Advanced Grant project 
``FLAVOUR''~(267104). Both M.F.\ and J.T.\ would like to thank DESY for
its hospitality. M.R.\ would like to thank the UC Irvine, where part of this 
work was done, for hospitality. We thank the Simons Center for Geometry and 
Physics in Stony Brook, the Center for Theoretical Underground Physics and 
Related Areas (CETUP* 2012) in South Dakota and the Gump station on Moorea for 
their hospitality and for partial support during the completion of this work. 

\appendix

\section{Details on lattices}
\label{app:details}

\subsection{Bravais types and form spaces}
\label{sec:bravaisform}

One can classify lattices by the symmetry groups they obey. This is the concept 
of Bravais equivalent lattices. In more detail, denote the symmetry group of 
some lattice $\Lambda$ as $G\subset\GL{n,\ZZ}$. Obviously, the point group 
$P \subset G$, is a subgroup of it. Now, if two 
lattices give rise to the same finite unimodular group $G$, we call them 
Bravais equivalent. This equivalence generates a finite number of Bravais 
types of lattices for every dimension $n$. They have been classified for 
dimensions up to six \cite{Plesken:1984}.

The interesting task would now be to decide which Bravais type a given lattice 
belongs to. This can be done using the notion of form spaces 
\cite{Plesken:1998}. The form space $\form{F}(G)$ of some finite group 
$G\subset\GL{n,\ZZ}$ is defined as the vector space of all symmetric 
matrices left invariant by $G$, i.e.\
\begin{equation}
\form{F}(G) ~=~ \{F~\in~\mathbbm{R}^{n\times n}_{\mathrm{sym}}~|~g^T\,F\,g~=~F \;\text{for all}\; g~\in~ G\}\;.
\label{eq4}
\end{equation}
On the other hand, we define the Gram matrix of the lattice basis 
$\basis{e}=\{e_1,\ldots,e_n\}$ as 
\begin{equation}
\gram(\basis{e})_{ij} ~=~ \scalarproduct{e_i}{e_j} ~=~ (\bmat_{\basis{e}}^{\,\mathrm{T}}\bmat_{\basis{e}})_{ij}\;,
\end{equation}
where the parentheses $\scalarproduct{e_i}{e_j}$ denote the standard scalar 
product. By definition, the Gram matrix is a symmetric, positive definite 
matrix. Under a change of lattice basis, represented by a unimodular matrix 
$M$, the Gram matrix changes as $M^T\gram(\basis{\basis{e}})\,M$, 
c.f.\ \Secref{sec:lattice}. By contrast, elements of the point group leave the 
Gram matrix invariant, i.e.\ for $\vartheta\in P$
\begin{equation}
\gram(\basis{e}) ~\stackrel{\vartheta}{~\longmapsto~}~ \vartheta^T\gram(\basis{\basis{e}})\,\vartheta ~=~ \gram(\basis{\basis{e}})\;. 
\end{equation}
Hence, a form space is in one--to--one correspondence to a Bravais type of 
lattice, i.e.\ every lattice $\Lambda$ has a basis 
$\basis{e}=\{e_1,\ldots,e_n\}$ such that its Gram matrix $\gram(\basis{e})$ is 
an element of the form space of a finite subgroup $P$ of $\GL{n,\ZZ}$, i.e.\ 
$\gram(\basis{e}) \in \form{F}(P)$ \cite{Brown:1978}. But in order to see 
that one lattice belongs to a given form space, it needs to be in this special 
basis, which is canonically chosen to be the shortest possible basis for that 
lattice. Fortunately, algorithms for precisely that task do exist, cf.\
e.g.\ 
\cite{Lenstra:1982} (though one should be careful: the shortest basis of a 
lattice is in general not unique).

Note that physically the Gram matrix is the metric of the torus defined by the 
lattice $\Lambda$ and the dimension of the form space $\form{F}(P)$  is exactly 
the number of (untwisted) moduli the orbifold offers.

Let us consider an example in two dimensions. Take the point group defined by 
\begin{equation}
P~=~\{\Id~=~\vartheta^2,\vartheta\}~\cong~\Z{2}\quad\mathrm{with}\quad
\vartheta~=~\left(\begin{array}{cr}1&0\\0&-1\end{array}\right)\;.
\end{equation}
It leaves invariant the form space
\begin{equation}
\form{F} (P)~=~\left(\begin{array}{cc}a^2&0\\0&b^2\end{array}\right)\;.
\end{equation}
That form space corresponds to the Bravais type called p--rectangular lattice 
(cf.\ \Apref{app:bravaislie}), consisting of two arbitrarily long, orthogonal 
vectors.

\subsection{Introducing an additional shift}
\label{app:latt:add}

DW~\cite{Donagi:2008xy} starts with an orthonormal lattice in 
six dimensions. Then, in a second step, additional shifts, which are linear 
combinations of the  (orthonormal) lattice vectors with rational coefficients, 
are included in the  space group. As we have seen in the second example in 
\Secref{subsec:class:ex},  those additional shifts can be incorporated to the 
lattice itself. Here we show  in detail how to transform the space group 
accordingly.

The perhaps most elegant procedure is to perform a change of basis, but using 
transformations from $\GL{n,\QQ}$. Hence, we are selecting a different 
\ZZ--class from the same \QQ--class, cf.\ \Secref{sec:class}. Let us list the 
necessary steps and illustrate them with an example:
\begin{enumerate}
\item The additional shift is a linear combination with rational coefficients 
of some of the lattice vectors. Exchange one of the old lattice 
vectors (that appears in the linear combination) by the new additional shift.
\item Write the transformation matrix $M$: start with the identity matrix and 
substitute the column corresponding to the exchanged vector by the coefficients 
of the linear combination.
\item Transform your space group using $M$ accordingly: see \Eqref{eq:changeb} 
and \Eqref{eqn:BasisChangeTheta}.
\item (Optional) In order to see the geometry more clearly, one can perform a 
basis reduction (e.g.\ using the LLL algorithm, cf.\ \cite{Lenstra:1982}), 
which is a transformation from $\GL{n,\ZZ}$.
\end{enumerate}

As an example, take the $\ZxZ{2}{2}$ model named (1--1) in
DW~\cite{Donagi:2008xy}, which consists of an orthogonal lattice
(p--cubic) with  orthonormal basis $\basis{e}$ and an additional shift
\begin{equation}
\tau ~=~ \frac{1}{2}\left(e_2 + e_4 + e_6 \right)\;.
\end{equation}
We will restrict the discussion to the three--dimensional (sub--)lattice 
$\Lambda$ spanned by the basis $\basis{e}=\{e_2,e_4,e_6\}$.

The basis matrix, Gram matrix and point group generators read
\begin{subequations}
\begin{align}
\bmat_{\basis{e}}&~=~\mthree{1&0&0\\0&1&0\\0&0&1}{2}\;,
&
\gram ({\basis{e}})&~=~\mthree{1&0&0\\0&1&0\\0&0&1}{2}\;,\\
\vartheta_{\basis{e}}&~=~\mthree{1&0&0\\0&-1&0\\0&0&-1}{2}\;,
&
\omega_{\basis{e}}&~=~\mthree{-1&0&0\\0&1&0\\0&0&-1}{2}\;.
\end{align}
\end{subequations}

Let us follow the steps described above:
\begin{enumerate}
\item We choose to exchange the $3^\mathrm{rd}$ (originally $6^\mathrm{th}$) vector for the additional
shift: the new basis $\basis{f}$ is spanned by $\basis{f}=\{e_2,e_4,\tau\}$. 
Notice that $\basis{f}$ is not a basis of the lattice $\Lambda$, but one of a
new, different lattice $\Sigma$.

\item In accordance with our choice, the transformation matrix is
\begin{equation}
M ~=~ \mthreer{1&0&\nicefrac{1}{2}\\0&1&\nicefrac{1}{2}\\0&0&\nicefrac{1}{2}}{2}\;.
\end{equation}
\item We perform the transformation using $M$. For the new lattice $\Sigma$ in 
the new basis $\basis{f}$, the quantities we are interested in look like
\begin{subequations}
\begin{align}
\bmat_{\basis{f}}&~=~\mthreer{1&0&\nicefrac{1}{2}\\0&1&\nicefrac{1}{2}\\0&0&\nicefrac{1}{2}}{2},
&
\gram({\basis{f}})&~=~ \mthreer{1&0&\nicefrac{1}{2}\\0&1&\nicefrac{1}{2}\\
\nicefrac{1}{2}&\nicefrac{1}{2}&\nicefrac{3}{4}}{2},\\
\vartheta_{\basis{f}}&~=~\mthreer{1&0&1\\0&-1&0\\0&0&-1}{2},
&
\omega_{\basis{f}}&~=~\mthreer{-1&0&0\\0&\phantom{-}1&1\\0&0&-1}{2}\;.
\end{align}
\end{subequations}

\item Next, we perform a LLL reduction, which is a change of
basis to a reduced one $\basis{r}$, and transform the point group elements
accordingly,
\begin{subequations}\label{eq:app:1-1}
\begin{align}
\bmat_{\basis{r}}&~=~\mthreer{
 \nicefrac{1}{2}& \nicefrac{1}{2}&-\nicefrac{1}{2}\\
 \nicefrac{1}{2}&-\nicefrac{1}{2}& \nicefrac{1}{2}\\
 \nicefrac{1}{2}&-\nicefrac{1}{2}&-\nicefrac{1}{2}}{2},
&
\gram ({\basis{r}})&~=~\frac{1}{4}\mthreer{3&-1&-1\\-1&3&-1\\-1&-1&3}{2}, \\
\vartheta_{\basis{r}}&~=~\mthreer{0&1&-1\\1&0&-1\\0&0&-1}{2},
&
\omega_{\basis{r}}&~=~\mthreer{0&-1&1\\0&-1&0\\1&-1&0}{2}\;.
\end{align}
\end{subequations}
\end{enumerate}

Last, we compare the Gram matrix $\gram ({\basis{r}})$ with 
\Tabref{tab:bravais-lie}. We see that introducing the additional shift $\tau$ 
into the p--cubic lattice is equivalent to work with the appropriately 
transformed point group in an i--cubic lattice.

A remark is in order. The form space left invariant by the $\ZxZ{2}{2}$ point 
group in the (reduced) basis of \Eqref{eq:app:1-1} is
\begin{equation}
\form{F} (P)~=~\left(\begin{array}{ccc}a&b&c\\b&a&-a-b-c\\c&-a-b-c&a\end{array}\right)\;.
\end{equation}
This form space is the one of a three--parametric, i--orthogonal lattice, which 
contains as possible realizations the i--cubic and the f--cubic lattices 
(both one--parametric, see table \ref{tab:bravais-lie}). Therefore, model 
(1--1) in \cite{Donagi:2008xy} corresponds to model $A_4$ 
of F{\"o}rste et al.\ \cite{Forste:2006wq}, i.e.\ to the Lie lattice 
$\SU{4}\times\SU{2}^3$ where the $\SU{4}$ part is an f--cubic lattice, see 
\Tabref{tab:z2z2}.

\subsection{Bravais types and Lie lattices}
\label{app:bravaislie}

It is common in the string--orbifold literature to describe lattices as root 
lattices of (semi--simple) Lie algebras. On the one hand, this makes it easy to 
identify the point group, i.e.\ a discrete subgroup of $\SU{3}$, using Weyl 
reflections and the Coxeter element. However, we find this practice to be 
problematic for at least three different reasons:

\subsubsection*{Redundancies}
\begin{figure}[ht]
\centering
\includegraphics[width=0.8\textwidth]{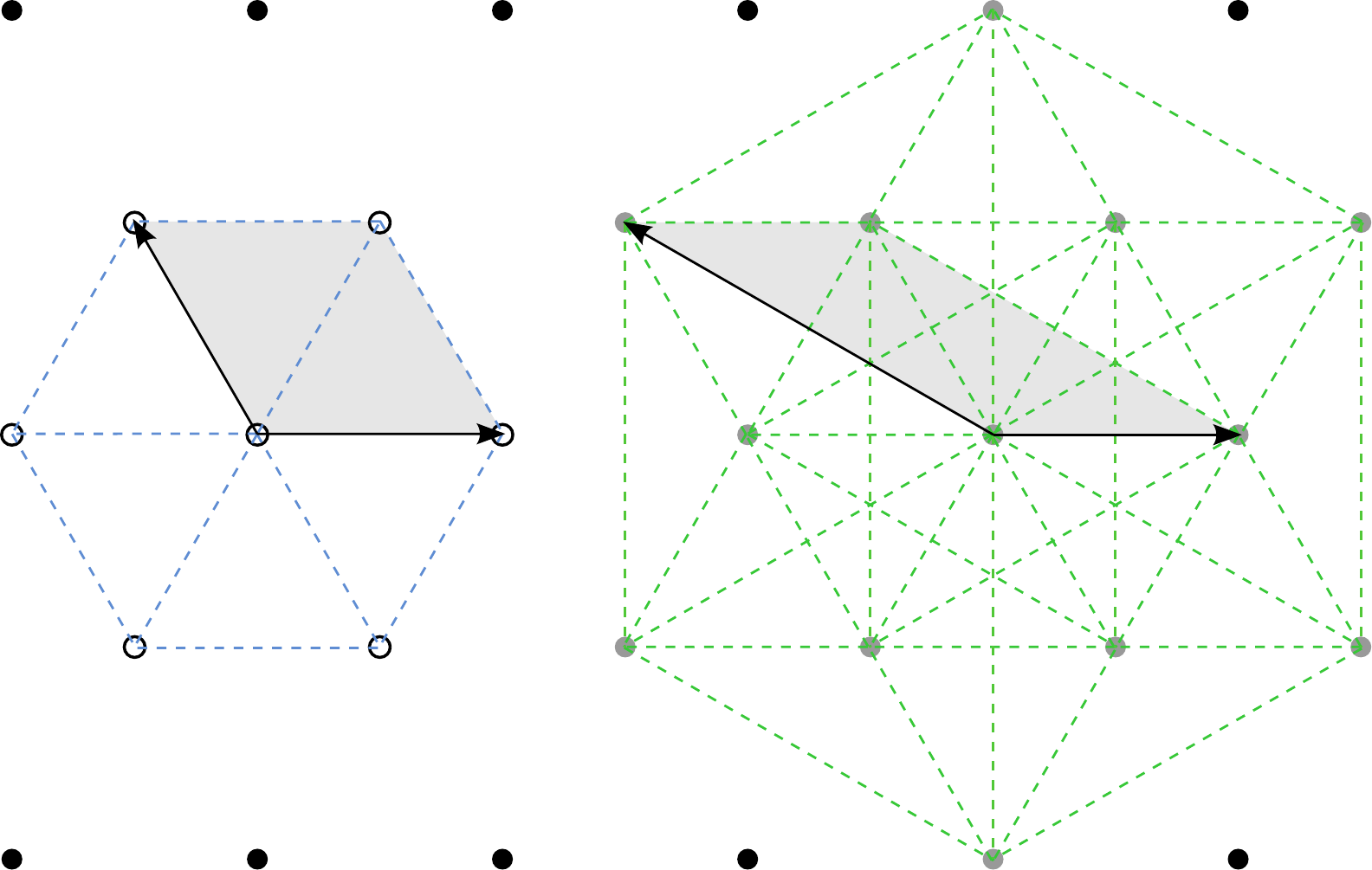}
\caption{The hexagonal lattice: the blue lines form the $\mathrm{SU}(3)$ root system, and the green lines form the $\mathrm{G}_2$ root system. Simple roots are also indicated, as well as the fundamental cells (shaded).}
\label{fig:su3g2}
\end{figure}

A root lattice is the lattice spanned by the simple roots of a certain 
(semi--simple) Lie algebra. Even if the simple roots of two non--equivalent 
(semi--simple) Lie algebras are different, the lattices they span might not. For 
example, the lattices spanned by the root systems of $\mathrm{SU}(3)$ and 
$\mathrm{G}_2$ are the same (see \Figref{fig:su3g2}). 
Some more examples are provided in \Tabref{tab:bravais-lie}.

\subsubsection*{Missing lattices}
When considering the redundancy of root lattices, one might think that there 
are more root lattices than types of lattices and that the situation could be 
resolved by introducing some clever convention to avoid this overcounting. But 
the problem exists in the other direction too: the set of all possible root 
lattices does not exhaust the whole family of Bravais types, i.e.\ there are 
Bravais types of lattices which are not generated by any root system. The 
lowest dimension in which this occurs is three and the most basic example is 
the body centered cubic lattice, also known as bcc or i--cubic to 
crystallographers (see \Tabref{tab:bravais-lie}). The bcc lattice is a cubic 
lattice with an additional lattice point in the center of the unit cell. Its 
only free parameter is the size of the system (e.g.\ the edge length of the 
cube). One possible way to convince oneself that there is no root lattice that 
can generate this Bravais lattice is taking every rank three root lattice 
and calculating which Bravais lattice it generates. We find that the i--cubic 
lattice has no description as root lattice (see \Tabref{tab:bravais-lie}).

\subsubsection*{Continuous parameters}
Every Bravais type allows for a set of continuous deformations which conserve 
its symmetries. Those deformations are encoded and made explicit in the form 
space that defines that particular Bravais type (cf.\ \Apref{sec:bravaisform}). 
The form space tells us how many deformation parameters one Bravais type allows 
for, and what is the effect of them (to change lengths of or angles between 
basis vectors). The realization of that freedom in the context of root lattices 
is very limited: lattices of Lie algebras allow for just one parameter, the 
size of the system; and if one includes semi--simple Lie algebras (direct 
products of simple ones), one can choose different sizes for different 
sublattices, but never the angles between vectors, which are fixed to a limited 
set of values. So, for example, a two--dimensional oblique lattice, in which 
the angle between the basis vectors is arbitrary, could never be unambiguously 
expressed in terms of Lie root lattices.

\paragraph{} 
In conclusion, the language of root lattices is incomplete and ambiguous, and 
is lacking geometrical insight with respect to the language of Bravais types 
and form spaces, which is, therefore, the one used in this paper.

Nevertheless, in order to justify some of the matchings between our 
classification of space groups and the ones already existing in the 
literature, we present in \Tabref{tab:bravais-lie} a classification of all of 
the Bravais types of lattices in 1, 2 and 3 dimensions, together with their 
equivalent root lattices, if there are any. There, in order to overcome the 
discussed ambiguities in the root lattice language, some conventions have been 
used:
\begin{itemize}
\item \orth~means orthogonal product. Unspecified products should be 
understood orthogonal.
\item \north~means free--angle product. The scalar product of the roots is 
indicated as a subindex. Notice that in the cases in which we have used this 
product there is actually no equivalent Lie lattice description: a 
non--orthogonal product of semi--simple Lie algebras is not a semi--simple Lie 
algebra. These possibilities are written in italics.
\item \arrowback~means a product with the leftmost factor.
\item Equal subindices mean equal length of the roots or equal scalar products.
\item A subindex in an algebra whose simple roots are of different length 
stands for the squared length of the shortest simple root, e.g.\ G$_{2,a}$ 
means that the shortest simple root of G$_2$ has length squared $a$.
\end{itemize}

{\centering
\begin{longtable}{|c|lr|c|}
\hline
Gram matrix & \multicolumn{2}{c|}{lattice name} & Lie algebra notation \\
\hline
\hline
\endhead
\hline
\multicolumn{4}{|r|}{continued \ldots}\\
\hline
\endfoot
\endlastfoot
\hline\multicolumn{4}{|c|}{\large 1 dimension}\\
\hline
$\left(\begin{array}{c}a\end{array}\right)$ & Ruler & r & SU(2)\\
\hline\hline\multicolumn{4}{|c|}{ \large 2 dimensions}\\
\hline
\gtwo{a&0\\&a} & Square & tp & SO(5), SU(2)$_a$\orth SU(2)$_a$\\
\hline
\gtwo{a&\pm a/2\\&a} & Hexagonal & hp& SU(3)$_a$, G$_{2,a}$\\
\hline
\gtwo{a&0\\&b} & p--Rectangular & op& SU(2)$_a$\orth SU(2)$_{b}$\\
\hline
\gtwo{a&b\\&a} & c--Rectangular & oc& {\em SU(2)$_a$\north[b]SU(2)$_{a}$}\\
\hline
\gtwo{a&c\\&b} & Oblique & mp& {\em SU(2)$_a$\north[c]SU(2)$_b$}\\
\hline\hline\multicolumn{4}{|c|}{ \large 3 dimensions}\\
\hline
\small{\gthree{a&0&0\\&a&0\\&&a}{3}} & p--Cubic & cP& SO(7), SU(2)$_a$\orth SU(2)$_a$\orth SU(2)$_a$\\
\hline
\small{\gthree{a& a/2 &\ a/2 \\&a & a/2 \\&&a}{3}} & f--Cubic& cF& SU(4), Sp(6)\\
\hline
\small{\gthree{a&- a/3 &- a/3 \\&a &- a/3 \\&&a}{3}} & i--Cubic & cI& {\small (none)}\\
\hline
\small{\gthree{a &\pm a/2 & 0\\&a&0 \\&& b}{3}} & p--Hexagonal & hP& $[$SU(3)$_a$ or G$_{2,a}]$\orth SU(2)$_b$\\
\hline
\small{\gthree{a &b &b \\&a &b \\&&a }{3}} & r--Hexagonal & hR& {\em SU(2)$_a$\north[b]SU(2)$_a$\north[b]SU(2)$_a$\north[b]\arrowback}\\
\hline
\small{\gthree{a &0 &0 \\&a &0 \\&&b }{3}} & p--Tetragonal & tP& $[$SU(2)$_a$\orth SU(2)$_a$ or SO(5)$]$\orth SU(2)$_b$\\
\hline
\small{\gthree{a+2b & -a& -b\\& a+2b& -b\\&&a+2b }{0}} & i--Tetragonal & tI& (no simple expr.)\\
\hline
\small{\gthree{a &0 &0 \\&b &0 \\&&c }{3}} & p--Orthorhombic & oP& SU(2)$_a$\orth SU(2)$_b$\orth SU(2)$_c$\\
\hline
\small{\gthree{a &c &0 \\&a &0 \\&&b }{3}} & c--Orthorhombic & oC& {\em SU(2)$_a$\north[c]SU(2)$_a$\orth SU(2)$_b$}\\
\hline
\small{\gthree{a+b &a &b \\&a+c &c \\&&b+c }{1}} & f--Orthorhombic & oF& (no simple expr.)\\
\hline
\small{\gthree{a+b+c &-a & -b\\&a+b+c & -c\\&&a+b+c }{0}} & i--Orthorhombic & oI& (no simple expr.)\\
\hline
\small{\gthree{a &c &0 \\&b &0 \\&&d }{3}} & p--Monoclinic & mP& {\em SU(2)$_a$\north[c]SU(2)$_b$\orth SU(2)$_d$}\\
\hline
\small{\gthree{a &c &d \\&a &d \\&& b}{3}} & c--Monoclinic & mC&  {\em SU(2)$_a$\north[c]SU(2)$_a$\north[d]SU(2)$_b$\north[d]\arrowback}\\
\hline
\small{\gthree{a &d &f \\&b &e \\&&c }{3}} & Triclinic & aP&  {\em SU(2)$_a$\north[d]SU(2)$_b$\north[e]SU(2)$_c$\north[f]\arrowback}\\
\hline
\caption{List of Bravais types in 1, 2 and 3 dimensions, together with possible 
root lattice expressions. The following prefixes and suffixes are used for the 
lattice names: {\em p} primitive, {\em c} centered (in 2D) or base--centered 
(in 3D), {\em f} face--centered, {\em i} body--centered, and {\em r} 
rhombohedral.
\label{tab:bravais-lie}}
\end{longtable}
}

In general, Bravais types with two or more parameters in the form space 
contain as specific cases other types with a lower number of parameters. For 
example, if we set the off diagonal parameter to zero in the two--dimensional 
oblique lattice (mp) (i.e.\ we take the basis vectors to be orthogonal), we 
get a p--rectangular (op) lattice. If we set now the diagonal elements of the 
form space to be equal (i.e.\ we take the basis vectors to have equal length), 
we get a square lattice (tp). These three lattices form the embedding chain 
tp\emb op\emb mp.

\begin{figure}[ht]
\centering
\begin{tabular}{cc}
\raisebox{0.4\height}{\includegraphics[width=0.38\textwidth]{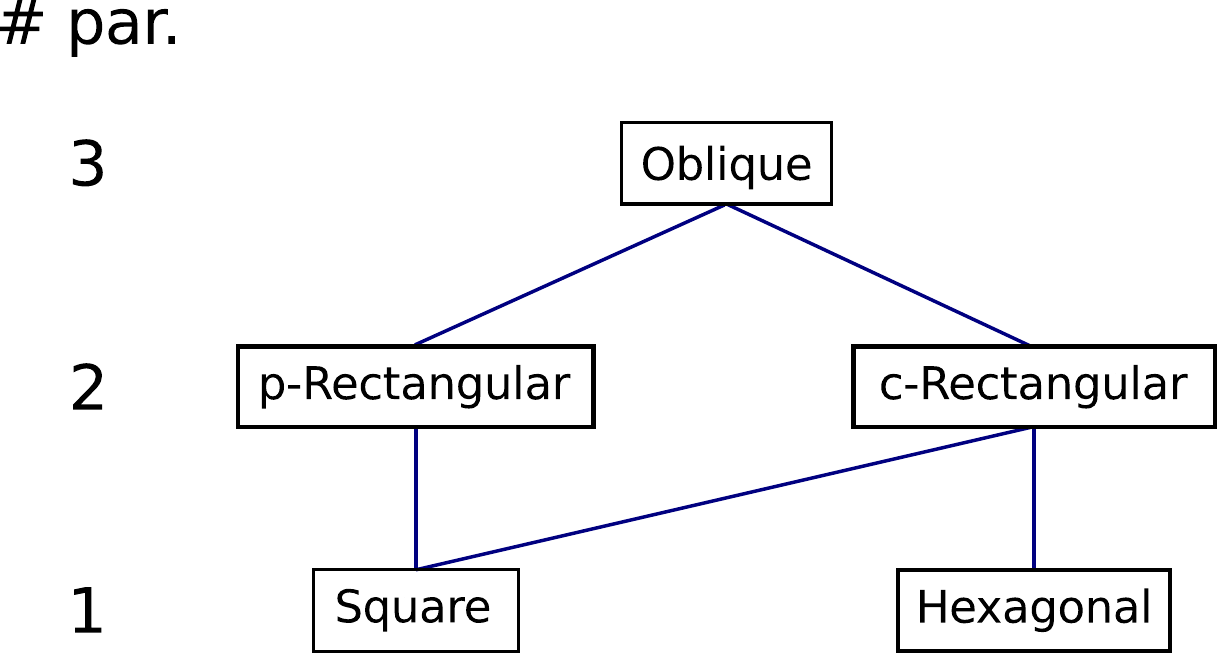}}&
\includegraphics[width=0.57\textwidth]{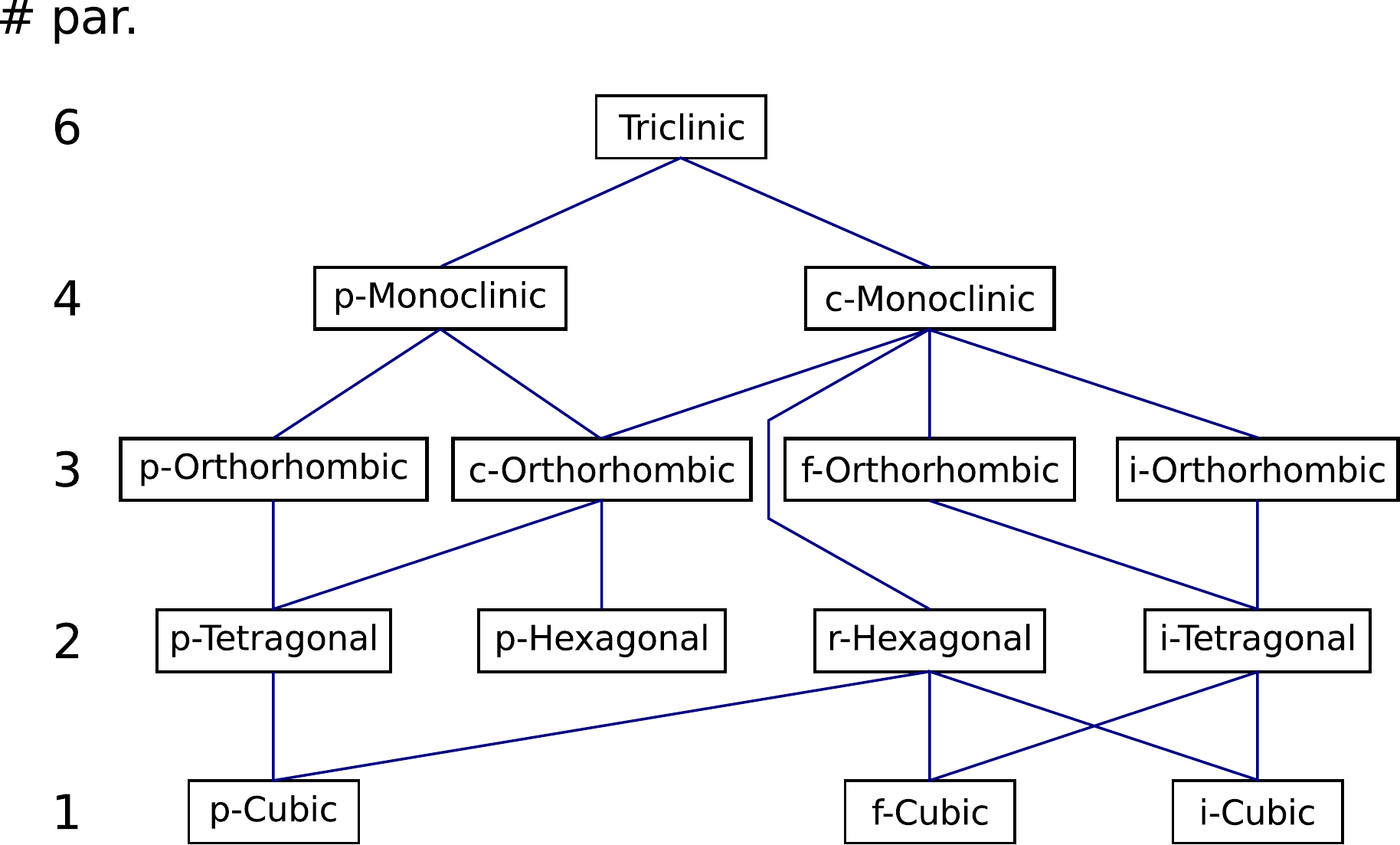}
\end{tabular}
\caption{Graph of Bravais types embeddings in 2D and 3D.}
\label{fig:emb}
\end{figure}

A graph containing all of the existing embeddings of that kind in two and three 
dimensions can be seen in \Figref{fig:emb}. For further information about this 
topic, the standard reference is \cite{Itc:2005}.

\section{Two--dimensional orbifolds}
\label{app:2dimorb}

In order to illustrate some of the concepts addressed in this paper, we 
reproduce here the list of all possible two--dimensional space groups, also 
known as {\em wallpaper groups}. They are well--known, and their classification 
can be found for instance in \cite{Brown:1978}.

The possible orders $m$ of point group elements in $n$ dimensions are given by 
the equation
\begin{equation}
\phi(m)~\le~ n\;,
\end{equation}
where $\phi$ is the Euler $\phi$--function. For dimension two, this leaves only
elements with order in $\{1,2,3,4,6\}$ as possible point group elements. In six
dimensions, this gets extended to $\{1, 2, 3,$ $4, 5, 6,$ $7, 8, 9,$ $10, 12, 14, 
18\}$. Nevertheless, in dimensions $n\ge2$, one can find point group elements 
with order $m$ such that $\phi(m)>n$. This can be realized by building a point 
group element as the direct sum of two point group elements of dimensions 
that add up to $n$. In that case, the order of the point group element would 
obviously be the least common multiple of the orders of the factors. For 
example, in six dimensions there exist point groups with elements of order 30, 
which are a direct sum of a four--dimensional order 10 element and a 
two--dimensional order 3 element.

\begin{table}[ht]
\centering
\begin{tabular}{|l|c|c|} \hline
\multicolumn{1}{|c|}{label of}   & \# of        & \# of affine \\
\multicolumn{1}{|c|}{\QQ--class} & \ZZ--classes & classes \\
\hline
\hline
$\mathrm{id}$                           & 1 & 1 \\
$\Z{2}$--I                              & 1 & 1 \\
$\Z{2}$--II                             & 2 & 3 \\
$\ZxZ{2}{2}\;\cong D_2$                 & 2 & 4 \\
$\Z{4}$                                 & 1 & 1 \\
$\Z{2}\ltimes\Z{4}\;\cong D_4$          & 2 & 2 \\
$\Z{3}$                                 & 1 & 1 \\
$\Z{2}\ltimes\Z{3}\;\cong S_3\cong D_3$ & 2 & 2 \\
$\Z{6}$                                 & 1 & 1 \\
$\Z{2}\ltimes\Z{6}\;\cong D_6$          & 1 & 1 \\
\hline
\end{tabular}
\caption{\QQ--classes in two dimensions.}
\label{tab:2dqclasses}
\end{table}

As discussed in \Secref{sec:class}, one can classify the $17$ two--dimensional 
space groups by their \QQ--classes. Those can be found in 
\Tabref{tab:2dqclasses}. There, $D_n$ is the dihedral group of order $2n$ 
and $S_n$ is the symmetric group of order $n!$. In \Tabref{tab:2daffclasses} 
the specific information of every affine class is shown: the \QQ--, \ZZ-- and 
affine class to which they belong, its Bravais type of lattice 
(cf.\ \Tabref{tab:bravais-lie}), its orbifolding group generators in augmented 
matrix notation and a name, description and image of its topology. The 
augmented matrix of some element 
$g_\basis{e}=(\vartheta_\basis{e}, \lambda_i e_i)\in S$ is given by
\begin{equation}
g_\basis{e} ~=~ \left(\begin{array}{c|c}\vartheta_\basis{e} & \lambda_i \\ \hline 0 & 1\end{array}\right)\;,
\end{equation}
using the lattice basis $\basis{e}$. This matrix acts on an augmented vector 
$(x,1)$ by simple matrix--vector multiplication.

\setlength\fboxsep{0.2cm}
\setlength\fboxrule{0pt}
\begin{longtable}{|p{2.7cm}|p{5.9cm}|p{4cm}|p{2.1cm}|} \hline
\QQ--\ZZ--aff. class\newline lattice & generators & name \& description & image\\ \hline \hline \endhead
\hline
\multicolumn{4}{|r|}{continued \ldots}\\
\hline
\endfoot
\endlastfoot
$\mathrm{id}$--1--1\vskip2mm Oblique 
& 
& Torus\vskip2mm Manifold
& \fbox{\raisebox{-0.75cm}{\includegraphics[width=1.7cm]{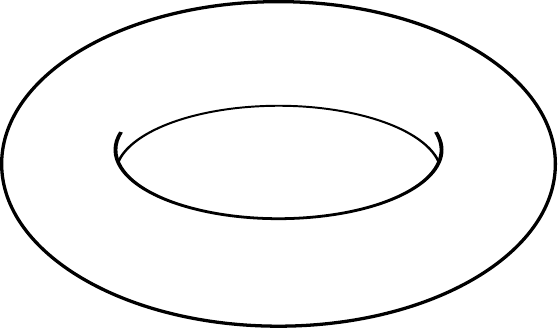}}}
\\ \hline \hline
$\Z{2}$--I--1--1\vskip2mm Oblique
& \raisebox{-0.55cm}{\small{$\left(\begin{array}{cc|c}-1 & 0 & 0 \\ 0 & -1 & 0 \\ \hline 0 & 0 & 1\end{array}\right)$}}
& Pillow\vskip2mm \small{Orbifold, 4 singularities with cone--angle~$\pi$}
& \fbox{\raisebox{-1cm}{\includegraphics[width=1.7cm]{pillow.pdf}}}
\\ \hline \hline
$\Z{2}$--II--1--1\vskip2mm\mbox{p--Rectangular}
& \raisebox{-0.5cm}{\small{$\left(\begin{array}{cc|c}1 & 0 & 0 \\ 0 & -1 & 0 \\ \hline 0 & 0 & 1\end{array}\right)$}}
& Pipe\vskip2mm Manifold, 2 boundaries
& \fbox{\raisebox{-0.8cm}{\includegraphics[width=1.7cm]{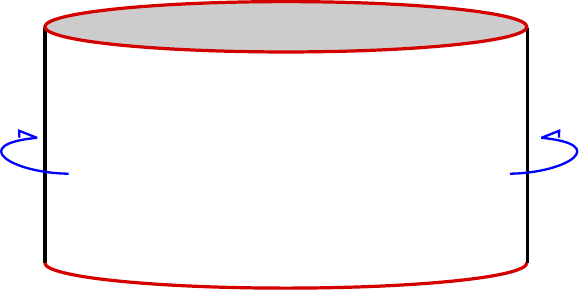}}}
\\ \hline
$\Z{2}$--II--1--2\vskip2mm \mbox{p--Rectangular}
& \raisebox{-0.5cm}{\small{$\left(\begin{array}{cc|c}1 & 0 & \nicefrac{1}{2} \\ 0 & -1 & 0 \\ \hline 0 & 0 & 1\end{array}\right)$}}
& Klein bottle\vskip2mm \small{Manifold, non--orientable}
& \hspace{2mm}\fbox{\raisebox{-1cm}{\includegraphics[width=1.3cm]{klein.pdf}}}
\\ \hline
$\Z{2}$--II--2--1\vskip2mm \mbox{c--Rectangular}
& \raisebox{-0.6cm}{\small{$\left(\begin{array}{cc|c}0 & 1 & 0 \\ 1 & 0 & 0 \\ \hline 0 & 0 & 1\end{array}\right)$}}
& M{\"o}bius strip\vskip2mm \small{Manifold, non--orientable, 1 boundary}
& \fbox{\raisebox{-1.2cm}{\includegraphics[width=1.7cm]{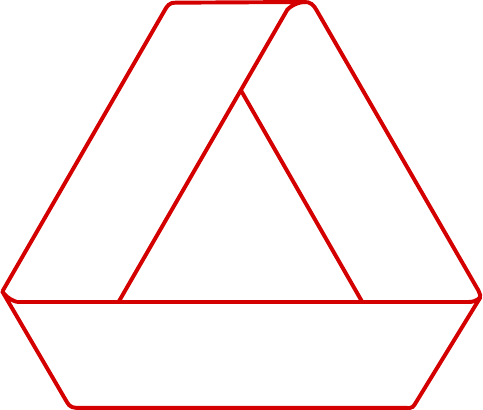}}}
\\ \hline \hline
$\ZZ_2\times\ZZ_2$--1--1\vskip2mm \mbox{p--Rectangular}
& \raisebox{-0.4cm}{\small{$\left(\begin{array}{cc|c}-1 & 0 & 0 \\ 0 & -1 & 0 \\ \hline 0 & 0 & 1\end{array}\right),\left(\begin{array}{cc|c}1 & 0 & 0 \\ 0 & -1 & 0 \\ \hline 0 & 0 & 1\end{array}\right)$}}
& Rectangle\vskip2mm Manifold, 1 boundary
& \fbox{\raisebox{-0.8cm}{\includegraphics[width=1.7cm]{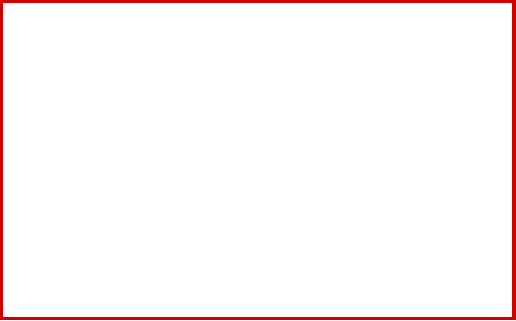}}}
\\ \hline
$\ZZ_2\times\ZZ_2$--1--2\vskip2mm \mbox{p--Rectangular}
& \raisebox{-0.8cm}{\small{$\left(\begin{array}{cc|c}-1 & 0 & 0 \\ 0 & -1 & 0 \\ \hline 0 & 0 & 1\end{array}\right),\left(\begin{array}{cc|c}1 & 0 & 0 \\ 0 & -1 & \nicefrac{1}{2} \\ \hline 0 & 0 & 1\end{array}\right)$}}
& Cut pillow\vskip2mm \small{Orbifold, 2 singularities with cone--angle $\pi$, 1 boundary}
& \fbox{\raisebox{-1.25cm}{\includegraphics[width=1.7cm]{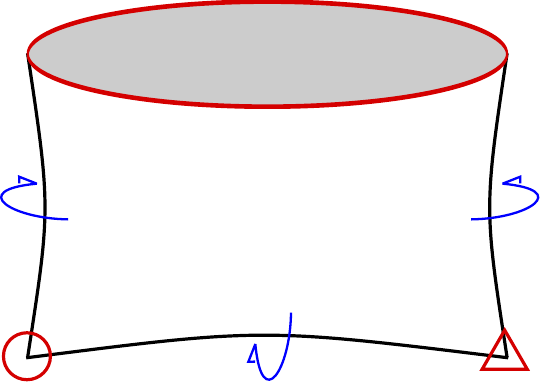}}}
\\ \hline
$\ZZ_2\times\ZZ_2$--1--3\vskip2mm \mbox{p--Rectangular}
& \raisebox{-0.4cm}{\small{$\left(\begin{array}{cc|c}-1 & 0 & 0 \\ 0 & -1 & 0 \\ \hline 0 & 0 & 1\end{array}\right),\left(\begin{array}{cc|c}1 & 0 & \nicefrac{1}{2} \\ 0 & -1 & \nicefrac{1}{2} \\ \hline 0 & 0 & 1\end{array}\right)$}}
& Cross--cap pillow\vskip2mm \small{Orbifold, 2 singularities with cone--angle $\pi$} 
& \hspace{1.5mm}\fbox{\raisebox{-0.9cm}{\includegraphics[width=1.4cm]{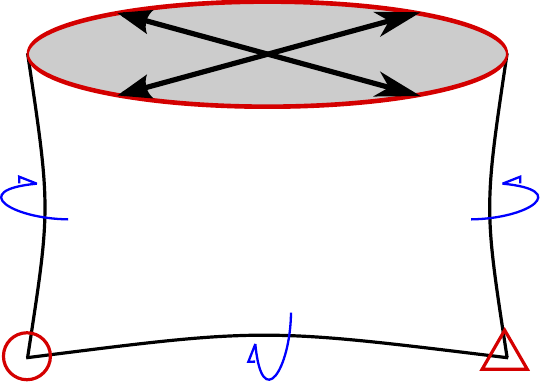}}}
\\ \hline
$\ZZ_2\times\ZZ_2$--2--1\vskip2mm \mbox{c--Rectangular}
& \raisebox{-0.8cm}{\small{$\left(\begin{array}{cc|c}-1 & 0 & 0 \\ 0 & -1 & 0 \\ \hline 0 & 0 & 1\end{array}\right),\left(\begin{array}{cc|c}0 & 1 & 0 \\ 1 & 0 & 0\\ \hline 0 & 0 & 1\end{array}\right)$}}
& Jester's hat\vskip2mm \small{Orbifold, 1 singularity with cone--angle~$\pi$, 1 boundary}
& \fbox{\raisebox{-1.25cm}{\includegraphics[width=1.5cm]{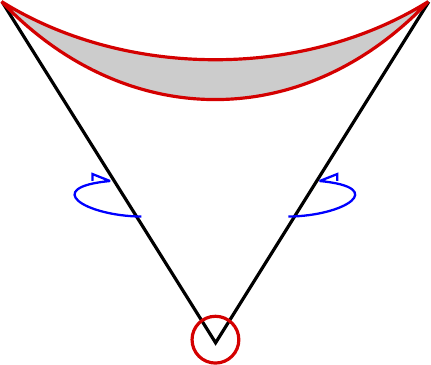}}}
\\ \hline \hline
$\ZZ_4$--1--1\vskip2mm Square
& \raisebox{-0.7cm}{\small{$\left(\begin{array}{cc|c}0 & -1 & 0 \\ 1 & 0 & 0 \\ \hline 0 & 0 & 1\end{array}\right)$}}
& Triangle pillow\vskip2mm \small{Orbifold, 2 singularities with cone--angle~$\nicefrac{\pi}{2}$, 1 singularity with cone--angle~$\pi$}
& \fbox{\raisebox{-1.2cm}{\includegraphics[width=1.7cm]{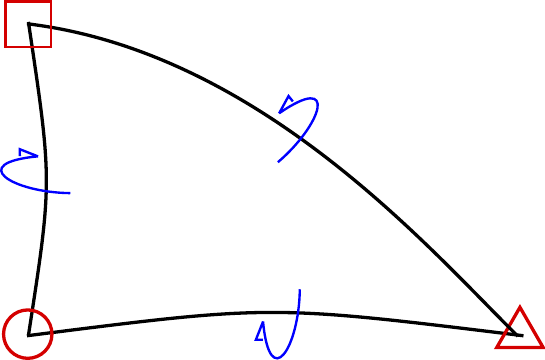}}}
\\ \hline \hline
$\ZZ_2\ltimes\ZZ_4$--1--1\vskip2mm Square
& \raisebox{-0.6cm}{\small{$\left(\begin{array}{cc|c}1 & 0 & 0 \\ 0 & -1 & 0 \\ \hline 0 & 0 & 1\end{array}\right),\left(\begin{array}{cc|c}0 & -1 & 0 \\ 1 & 0 & 0 \\ \hline 0 & 0 & 1\end{array}\right)$}}
& Triangle\vskip2mm \small{Manifold, one boundary, 1 angle of $\nicefrac{\pi}{2}$ and 2 of $\nicefrac{\pi}{4}$}
& \hspace{1mm}\fbox{\raisebox{-1.3cm}{\includegraphics[width=1.5cm]{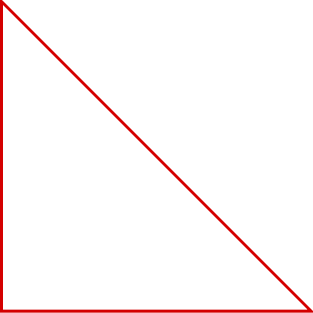}}}
\\ \hline
$\ZZ_2\ltimes\ZZ_4$--1--2\vskip2mm Square
& \raisebox{-0.7cm}{\small{$\left(\begin{array}{cc|c}1 & 0 & \nicefrac{1}{2} \\ 0 & -1 & \nicefrac{1}{2} \\ \hline 0 & 0 & 1\end{array}\right),\left(\begin{array}{cc|c}0 & -1 & 0 \\ 1 & 0 & 0 \\ \hline 0 & 0 & 1\end{array}\right)$}}
& Jester's hat\vskip2mm \small{Orbifold, 1 singularity with cone--angle~$\nicefrac{\pi}{2}$, 1 boundary}
& \hspace{1mm}\fbox{\raisebox{-1.25cm}{\includegraphics[width=1.5cm]{jhat.pdf}}}
\\ \hline \hline
$\ZZ_3$--1--1\vskip2mm \mbox{Hexagonal}
& \raisebox{-0.5cm}{\small{$\left(\begin{array}{cc|c}0 & -1 & 0 \\ 1 & -1 & 0 \\ \hline 0 & 0 & 1\end{array}\right)$}}
& Triangle pillow\vskip2mm \small{Orbifold, 3 singularities with cone--angle~$\nicefrac{2\pi}{3}$}
& \fbox{\raisebox{-1.cm}{\includegraphics[width=1.7cm]{tripillow.pdf}}}
\\ \hline \hline
$\ZZ_2\ltimes\ZZ_3$--1--1\vskip2mm \mbox{Hexagonal}
& \raisebox{-0.5cm}{\small{$\left(\begin{array}{cc|c}0 & -1 & 0 \\ -1 & 0 & 0 \\ \hline 0 & 0 & 1\end{array}\right),\left(\begin{array}{cc|c}0 & -1 & 0 \\ 1 & -1 & 0 \\ \hline 0 & 0 & 1\end{array}\right)$}}
& Triangle\vskip2mm \small{Manifold, 3 boundary, all angles $\nicefrac{\pi}{3}$}
& \hspace{1mm}\fbox{\raisebox{-1cm}{\includegraphics[width=1.5cm]{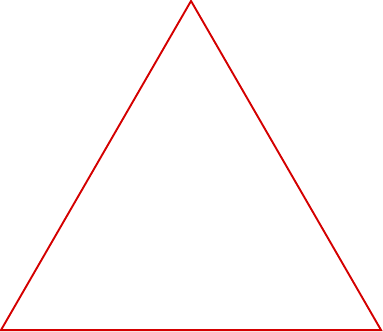}}}
\\ \hline
$\ZZ_2\ltimes\ZZ_3$--2--1\vskip2mm \mbox{Hexagonal}
& \raisebox{-0.7cm}{\small{$\left(\begin{array}{cc|c}0 & 1 & 0 \\ 1 & 0 & 0 \\ \hline 0 & 0 & 1\end{array}\right),\left(\begin{array}{cc|c}0 & -1 & 0 \\ 1 & -1 & 0 \\ \hline 0 & 0 & 1\end{array}\right)$}}
& Jester's hat\vskip2mm \small{Orbifold, 1 singularity with cone--angle~$\nicefrac{2\pi}{3}$, 1 boundary}
& \hspace{1mm}\fbox{\raisebox{-1.25cm}{\includegraphics[width=1.5cm]{jhat.pdf}}}
\\ \hline \hline
$\ZZ_6$--1--1\vskip2mm \mbox{Hexagonal}
& \raisebox{-0.7cm}{\small{$\left(\begin{array}{cc|c}1 & -1 & 0 \\ 1 & 0 & 0 \\ \hline 0 & 0 & 1\end{array}\right)$}}
& Triangle pillow\vskip2mm \small{Orbifold, 3 singularities with cone--angles $\nicefrac{2\pi}{3}$, $\nicefrac{\pi}{3}$ and $\pi$}
& \fbox{\raisebox{-1.1cm}{\includegraphics[width=1.7cm]{tripillow.pdf}}}
\\ \hline \hline
$\ZZ_2\ltimes\ZZ_6$--1--1\vskip2mm \mbox{Hexagonal}
& \raisebox{-0.7cm}{\small{$\left(\begin{array}{cc|c}0 & 1 & 0\\ 1 & 0 & 0\\ \hline 0 & 0 & 1\end{array}\right),\left(\begin{array}{cc|c}1 & -1 & 0 \\ 1 & 0 & 0 \\ \hline 0 & 0 & 1\end{array}\right)$}}
& Triangle\vskip2mm \small{Manifold, 1 boundary, with angles $\nicefrac{\pi}{2}$, $\nicefrac{\pi}{3}$ and $\nicefrac{\pi}{6}$}
& \fbox{\raisebox{-1.2cm}{\includegraphics[width=1.7cm]{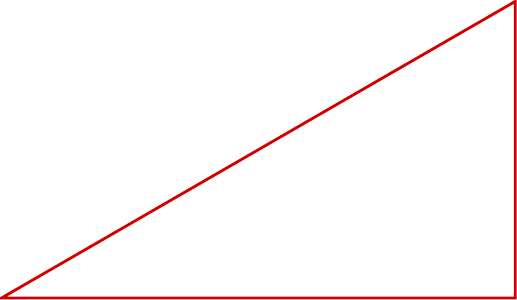}}}
\\ \hline
\caption{List of all possible two--dimensional orbifolds. \mbox{\QQ--classes} are separated by double lines.\label{tab:2daffclasses}}
\end{longtable}

Sometimes it is of interest to know the fundamental groups of the resulting 
orbifolds. Among the two--dimensional space groups, most of the fundamental 
groups are trivial with the following exceptions: the torus has a fundamental 
group of $(\ZZ)^2$, the pipe and the M{\"o}bius strip \ZZ, the cross--cap pillow (a projective 
plane) $\ZZ_2$ and the Klein bottle's one is its own space group, with group 
structure
\begin{equation}
S~=~\left\{a^nb^m\;|\;m,n\in\ZZ\,,\;b\,a~=~a^{-1}\,b\right\}\;.
\end{equation}

\newpage
\section{Tables}
\label{app:results}

\subsection{Abelian point groups}

\begin{center}
\begin{longtable}{|c|c|c|l|c|}
\hline
\QQ--		& \ZZ--		& affine   & generators of $G$ & \\
\!\!class\!\!	& \!\!class\!\!	& class,   &                   & \\
\cline{4-4}
($P$)		& ($\Lambda$)	& breaking & contributions to $(h^{(1,1)},h^{(2,1)})$ from $U$ and $T$ sectors & \!\!\!$(h^{(1,1)},h^{(2,1)})$\!\!\! \\
\hline
\hline
\endfirsthead
\hline
\QQ--		& \ZZ--		& affine   & generators of $G$ & \\
\!\!class\!\!	& \!\!class\!\!	& class,   &                   & \\
\cline{4-4}
($P$)		& ($\Lambda$)	& breaking & contributions to $(h^{(1,1)},h^{(2,1)})$ from $U$ and $T$ sectors & \!\!\!$(h^{(1,1)},h^{(2,1)})$\!\!\! \\
\hline
\hline
\endhead
\hline
\multicolumn{5}{|r|}{continued \ldots}\\
\hline
\endfoot
\endlastfoot
$\Z{3}$		&           1 &         1 & $\left(\theta, 0\right)$ 							& \\
\cline{4-4}
                &             &     local & $(9,0)U+(27,0)T_1$								& $(36,0)$\\
\hline
\hline
$\Z{4}$		&           1 &         1 & $\left(\theta, 0\right)$ 							& \\
\cline{4-4}
                &             &     local & $(5,1)U+(16,0)T_1+(10,6)T_2$						& $(31,7)$\\
\cline{2-5}
		&           2 &         1 & $\left(\theta, 0\right)$							& \\
\cline{4-4}
                &             &     local & $(5,1)U+(16,0)T_1+(6,2)T_2$							& $(27,3)$\\
\cline{2-5}
		&           3 &         1 & $\left(\theta, 0\right)$		 					& \\
\cline{4-4}
                &             &     local & $(5,1)U+(16,0)T_1+(4,0)T_2$							& $(25,1)$\\
\end{longtable}
\end{center}

\begin{figure}[ht]
\centering
\includegraphics[width=0.7\textwidth]{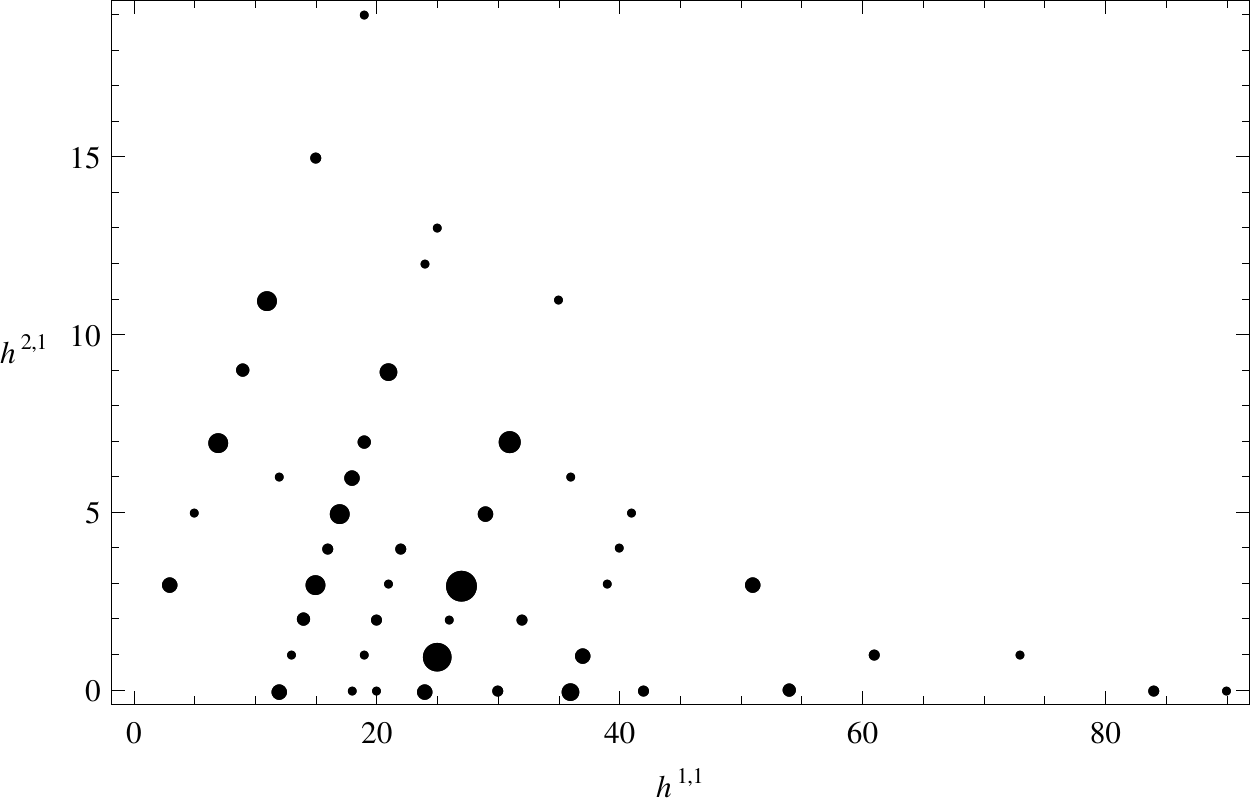}
\caption{Statistics of the Hodge numbers for the 138 Abelian toroidal orbifolds of \Tabref{tab:complist}.}
\label{fig:hodgenumbers}
\end{figure}

\begin{landscape}

\subsection[Non--Abelian point groups]{%
\texorpdfstring{Non--Abelian point groups}{Non-Abelian point groups}}

\begin{center}
\begin{longtable}{|c|c|l|c|}
\hline
label of \QQ--class & \Carat & twists        & $\#$ of conj.\\
GAPID               & index & from $\SU{3}$ & classes \\
\hline
\hline
\endfirsthead
\hline
label of \QQ--class & \Carat & twists        & $\#$ of conj.\\
GAPID               & index & from $\SU{3}$ & classes \\
\hline
\hline
\endhead
\hline
\multicolumn{4}{|r|}{continued \ldots}\\
\hline
\endfoot
\endlastfoot
\GLabels{S_3}{6,1} & 2262 &
$\left(
\begin{array}{ccc}
-1 & 0 & 0 \\
 0 & 0 & 1 \\
 0 & 1 & 0
\end{array}
\right)\;,
\left(
\begin{array}{ccc}
 1 & 0 & 0 \\
 0 & \mathrm{e}^{-2\pi\,\I\,\frac{1}{3}} & 0 \\
 0 & 0 & \mathrm{e}^{2\pi\,\I\,\frac{1}{3}}
\end{array}
\right)$ & 3 \\
\hline 
\GLabels{D_4}{8,3} & 4682 &
$\left(
\begin{array}{ccc}
-1 & 0 & 0 \\
 0 & 0 & 1 \\
 0 & 1 & 0
\end{array}
\right)\;,
\left(
\begin{array}{ccc}
-1 &  0 & 0 \\
 0 & -1 & 0 \\
 0 &  0 & 1
\end{array}
\right)$ & 5 \\
\hline 
\GLabels{Q_8\; (\mathcal{N}=2)}{8,4} & 5750 &
$\left(
\begin{array}{ccc}
 1 & 0 & 0 \\
 0 & 0 & 1 \\
 0 & -1 & 0
\end{array}
\right)\;,
\left(
\begin{array}{ccc}
 1 & 0 & 0 \\
 0 & -\I & 0 \\
 0 & 0 & \I
\end{array}
\right)$ & 5 \\
\hline 
\GLabels{\Dic{3}\; (\mathcal{N}=2)}{12,1} & 3374 & 
$\left(
\begin{array}{ccc}
 1 &  0 & 0 \\
 0 &  0 & 1 \\
 0 & -1 & 0
\end{array}
\right)\;,
\left(
\begin{array}{ccc}
 1 & 0 & 0 \\
 0 & \mathrm{e}^{-2\pi\,\I\,\frac{1}{3}} & 0 \\
 0 & 0 & \mathrm{e}^{-2\pi\,\I\,\frac{1}{3}}
\end{array}
\right)$ & 6 \\
\hline 
\GLabels{A_4}{12,3} & 4893 & 
$\left(
\begin{array}{ccc}
 0 & 1 & 0 \\
 0 & 0 & 1 \\
 1 & 0 & 0
\end{array}
\right)\;,
\left(
\begin{array}{ccc}
-1 & 0 & 0 \\
 0 & 1 & 0 \\
 0 & 0 & -1
\end{array}
\right)$ & 4 \\
\hline 
\GLabels{D_6}{12,4} & 2258 &
$\left(
\begin{array}{ccc}
-1 & 0 & 0 \\
 0 & 0 & 1 \\
 0 & 1 & 0
\end{array}
\right)\;,
\left(
\begin{array}{ccc}
 1 & 0 & 0 \\
 0 & \mathrm{e}^{2\pi\,\I\,\frac{1}{6}} & 0 \\
 0 & 0 & \mathrm{e}^{-2\pi\,\I\,\frac{1}{6}}
\end{array}
\right)$ & 6 \\
\end{longtable}
\end{center}

\end{landscape}

\bibliography{Orbifold}
\bibliographystyle{NewArXiv.bst}
\end{document}